\documentclass[onecolumn, numberedappendix]{aastex63}

\newcommand{\av}{$A_V$}

\newcommand{\etal}{et~al.}

\newcommand{\ks}{$K_{\rm s}$}
\newcommand{\vmkz}{$(V-K_{\rm s})_0$}
\newcommand{\vmk}{$(V-K_{\rm s})$}

\newcommand{\vsini}{$v \sin i$}
\newcommand{\mum}{$\mu$m}
\newcommand{\rop}{$\rho$ Oph}

\begin{document}

\title{Rotation of Low-Mass Stars in Taurus with K2 }


\author{L.~M.~Rebull}
\affiliation{Infrared Science Archive (IRSA), IPAC, 1200 E.\
California Blvd., California Institute of Technology, Pasadena, CA
91125, USA; rebull@ipac.caltech.edu}
\author{J.~R.~Stauffer}
\affiliation{Spitzer Science Center (SSC), (IPAC), 1200 E.\ California
Blvd., California Institute of Technology, Pasadena, CA 9112, USA}
\author{A.~M.~Cody}
\affiliation{Bay Area Environmental Research Institute, 625 2nd
St.\ Ste.\ 209, Petaluma, CA 94952, USA}
\author{L.~A.~Hillenbrand}
\affiliation{Astronomy Department, California Institute of
Technology, Pasadena, CA 91125, USA}
\author{J.~Bouvier}
\affiliation{Universit\'e de Grenoble, Institut de Plan\'etologie
et d'Astrophysique de Grenoble (IPAG), F-38000 Grenoble, France;
CNRS, IPAG, F-38000 Grenoble, France}
\author{N.~Roggero}
\affiliation{Universit\'e de Grenoble, Institut de Plan\'etologie
et d'Astrophysique de Grenoble (IPAG), F-38000 Grenoble, France;
CNRS, IPAG, F-38000 Grenoble, France}
\author{T.\ J.\ David}
\affiliation{Center for Computational Astrophysics, Flatiron
Institute, New York, NY 10010, USA}

\begin{abstract}  

We present an analysis of K2 light curves (LCs) from Campaigns
4 and 13 for members of the young ($\sim$3 Myr) Taurus association,
in addition to an older ($\sim$30 Myr) population of stars that is
largely in the foreground of the Taurus molecular clouds.  Out of 156
of the highest-confidence Taurus members, we find that 81\% are
periodic. Our sample of young foreground stars is biased and
incomplete, but nearly all (37/38) are periodic.  The overall
distribution of rotation rates as a function of color (a proxy for
mass) is similar to that found in other clusters: the slowest rotators
are among the early M spectral types, with faster rotation towards
both earlier FGK and later M types. The relationship between period
and color/mass exhibited by older clusters such as the Pleiades is
already in place by Taurus age. The foreground population has very few
stars, but is consistent with the USco and Pleiades period
distributions.  As found in other young clusters, stars with disks
rotate on average slower, and few with disks are found rotating faster
than $\sim$2 d. The overall amplitude of the light curves decreases
with age and higher mass stars have generally lower amplitudes than
lower mass stars. Stars with disks have on average larger amplitudes
than stars without disks, though the physical mechanisms driving the
variability and the  resulting light curve morphologies are also
different between these two classes.

\end{abstract}

\section{Introduction}
\label{sec:intro}

The Taurus-Auriga star forming region has fundamentally shaped our
understanding of how low mass stars form (see, e.g., Kenyon \&
Hartmann 1995). At an age of $\lesssim$3 Myr (e.g., Kraus \&
Hillenbrand 2009, Luhman 2018) and a distance of $\sim$140 pc (e.g.,
Esplin \& Luhman 2017), even the low-mass objects (mid-M to early L)
are bright enough to be well-studied. A significant fraction of the
members have infrared (IR) excesses indicative of circumstellar
disks.

From the earliest studies of the prototype T~Tauri and other stars in
Taurus, variability was included as a defining characteristic of young
stars (Joy 1945). The variability arises from starspots, flares, disk
interactions, disk accretion, disk structure occulting the star, and
more.  NASA's K2 mission (Howell \etal\ 2014) has recently provided
high quality, long duration ($\sim$70 d), high cadence ($\sim$30 min)
light curves (LCs) for stars in many different clusters. In two of
K2's Campaigns (4 and 13), members of the Taurus-Auriga molecular
cloud population were included. Those Taurus members, along with the
$\rho$ Oph members observed in K2's Campaign 2 (e.g., Rebull \etal\
2018) are the youngest cluster stars observed with K2.

Aperiodic variability is expected and common in young stars (see,
e.g., Cody \etal\ 2016, Cody \& Hillenbrand 2018). However, when the
variability is periodic, we can infer the rotation rate of the star,
and hence, we can explore the rotation rates of young stars as a
function of stellar mass and disk presence. When combined with
comparable observations of other clusters, we can study trends as a
function of age. In this paper, we explore the periodic variability of
the Taurus members observed with K2. Cody \etal\ (2020 in prep) will
explore the range of light curve (LC) properties of the disked
members.

We have previously published our analysis of the K2 data for the 
Pleiades ($\sim$125 Myr; Rebull \etal\ 2016a,b, Stauffer
\etal\ 2016b; papers I, II, and III, respectively), Praesepe
($\sim$790 Myr; Rebull \etal\ 2017; paper IV), and Upper
Sco/$\rho$ Oph ($\sim$8 Myr and $\sim$1 Myr; Rebull \etal\
2018; paper V). We have deliberately performed our analyses of these
clusters, now including Taurus, in a very homogeneous fashion in order
to best allow intercomparison of the rotation data across the full age
range observed with K2. Upper Sco has less reddening and fewer disks
than Taurus; $\rho$ Oph has more reddening than, and a comparable disk
fraction as, Taurus. While there are nearly 1000 member stars with
periodic LCs in each of the Pleiades, Praesepe, and Upper Sco, there
are many fewer stars with suitable K2 data at the youngest ages.  In
$\rho$ Oph, there are 174 member LCs, 106 ($\sim$60\%) of which are
periodic. There are comparable numbers in Taurus, where there are 156
highest quality members (see \S\ref{sec:membership} and
Table~\ref{tab:summarystats} below), and 81\% (127) of those are
periodic. In context with the older clusters, Taurus and $\rho$ Oph
seem to suffer in comparison, because of fewer stars, more stochastic
contributions to the LCs (from disks and accretion, generally yielding
a lower fraction of periodic LCs), more reddening and spatial
variability in reddening (affecting the scatter in the diagrams), and
more uncertainty in membership (yielding a higher contamination
rate). 

In Section~\ref{sec:obs}, we summarize the data we amassed, including
information about the K2 data, literature information, member
selection, dereddening, and disk identification.
Section~\ref{sec:interp} begins with period identification and
interpretation, and comparison of our periods to those from the
literature. This section ends with color-magnitude diagrams for the
sample.  Section~\ref{sec:disks} discusses the influence of disks on
the period distribution. Section~\ref{sec:rotationdistrib} presents
the distributions of periods and periods against color as a proxy for
mass. We also compare Taurus to the rest of the clusters we have
analyzed with K2 data (papers I-V).  In Section~\ref{sec:linkage}, we
characterize the LCs in the same fashion as we did for the other
clusters.  Finally, we summarize our results in
Section~\ref{sec:concl}.   

We note explicitly that there are four sets of stars discussed in this
paper: (1) highest quality Taurus members, (2) lower-quality members
(in other words, possible Taurus members), (3) a population largely
foreground to Taurus that is likely older than Taurus but still young
($\sim$30 Myr), and (4) non-members (NM).
Appendix~\ref{app:membership}  contains a detailed description
of how we define these categories, with an overview in
Section~\ref{sec:membership}. When we use the term ``members,'' we
mean the highest quality plus the possible Taurus members. When we use
the term ``foreground'' (as in ``foreground population''), we mean
this set of stars that are older than Taurus but still relatively
young.

\section{Data}
\label{sec:obs}

\subsection{K2 Data}
\label{sec:k2data}

\begin{figure}[ht]
\epsscale{1.0}
\plotone{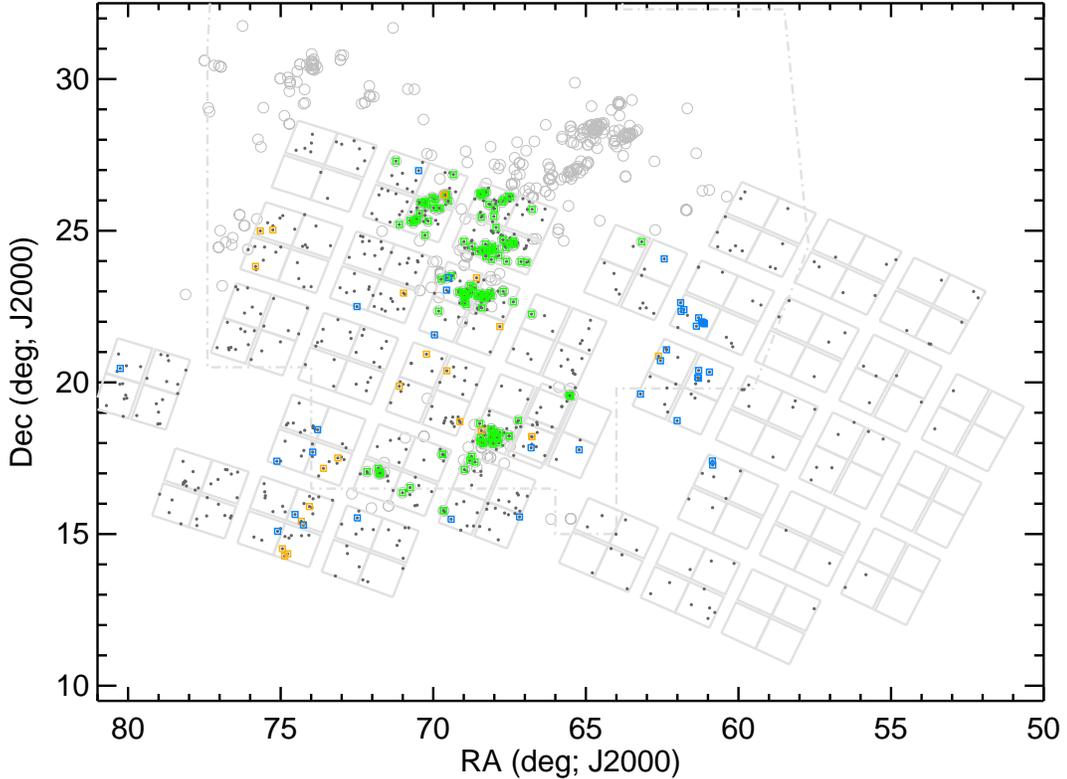}
\caption{Locations of targets of note projected onto the sky.  Small
dark grey circles: candidate Taurus members with LCs. Additional green
square: highest quality Taurus members (see
Sec.~\ref{sec:membership} for more on membership). Additional orange:
possible Taurus members. Additional blue square: foreground
population, e.g., stars determined in the literature to be young, but
not kinematically consistent with Taurus membership; these could be
members of, e.g., Group 29 (see Sec.~\ref{sec:membership}).  For
context, the large light grey squares are the K2 footprints (C13,
left; C4, right), and the small light grey circles are the Taurus
members as compiled by Esplin \& Luhman (2019); the dash-dot line is
the polygon considered for potential Taurus members using WISE in
Rebull \etal\ (2011). The K2 observations cover about 30\% of the 
likely Taurus members from Esplin \& Luhman (2019).}
\label{fig:where}
\end{figure}

Stars in the Taurus Molecular Cloud were observed in two different K2
campaigns, Campaign 4 (C4; 2015 Feb 7 -- 2015 Apr 24) and Campaign 13
(C13; 2017 Mar 8 -- 2017 May 27), with the majority of Taurus members
coming from C13; see Figure~\ref{fig:where}. As in our earlier papers,
we start by considering all possible cluster members and then narrow
the sample to high-confidence and lower confidence members (see
Sec.~\ref{sec:membership}); results for the likely NM are listed
in the Appendix for reference. There are $\sim$850 candidate Taurus
members with K2 LCs from either campaign and a few with K2 data from
both campaigns (see Table~\ref{tab:summarystats}).  All of the LCs
used here were observed in the long-cadence ($\sim$30 min cadence)
mode.

Our analysis (see Sec.~\ref{sec:membership} below) focused on the 156
highest quality Taurus members, the 23 possible members, and the 38 
foreground objects that are young (but likely older than the Taurus
molecular cloud population).  Figure~\ref{fig:where} shows the
distribution of these objects with K2 LCs on the sky, in context with
other reference points. The area defined as encompassing `Taurus' by,
e.g., Esplin \& Luhman (2019) goes further north (and east) than the
K2 region, such that the K2 observations cover about 30\% of the
members from Esplin \& Luhman (2019). Assuming that there should be no
spatial dependence of stellar rotation rate on location in the
cluster, there should be no bias in our distribution of periods
resulting from the incomplete coverage of the cluster.  However,
specifically because the K2 region extends over a larger area to the
south and west, it encompasses a larger area than has been studied
most intensively for potential Taurus members, and enables LC
collection from a more dispersed population of candidate young
stars.

There are very few stars considered to be Taurus members of any
confidence level that have LCs in both the C4 and C13 campaigns. Of
these, just  two are in the set of possible Taurus members (EPIC
210689309=AG+18339; EPIC 210689130=HD 28150), and one is in the
foreground population (210662824=HD 285778). All three of these are
periodic in both campaigns. There are an additional ten field stars
with two sets of LCs, and four of them have periods in both campaigns. 
For all the stars in common between C4 and
C13 (whether or not they are members), the periods derived separately
from both campaigns are the same within $\sim$4\% accuracy,
suggesting that the effective astrophysical uncertainty on the periods
is likely very low, at least for stars without disks.

Our analysis follows the same approach discussed in our earlier papers
(papers I-V), and so is briefly summarized here.  For each target, we
selected the best (see Paper I) LC from several different available LC versions:
(1)  The pre-search data conditioning (PDC) version generated by the
Kepler project\footnote{https://keplerscience.arc.nasa.gov/k2-pipeline-release-notes.html}
and obtained from MAST, the Mikulski Archive for Space
Telescopes. (2) A version using custom software developed by co-author
A.~M.~Cody. (3) The `self-flat-fielding' approach used by Vanderburg
\& Johnson (2014) and the K2SFF pipeline as obtained from MAST. (4)
The LCs from the EVEREST2 pipeline (Luger \etal\ 2016, 2017), which
uses pixel level decorrelation, as obtained from MAST. (5) The
EVEREST2 pipeline LCs, where large-scale structure is removed via a
least-squares (Savitzky-Golay) polynomial smoothing filter (e.g.,
Press \etal\ 1992), as implemented in the IDLastro
library\footnote{https://idlastro.gsfc.nasa.gov}.   We removed any
data points corresponding to thruster firings and any others with bad
data flags set in the corresponding data product.  Any periodic
signals are generally unambiguous, and are generally detected in all
the LC versions.  The spacecraft slowly drifts and then repositions
regularly every 0.245 d, so we particularly scrutinized any detected
periods near 0.245 d.

K2 has a relatively large pixel size ($3.98\arcsec \times 3.98
\arcsec$).  To identify sources subject to confusion, we inspected the
region around each target with the  Finder Chart
tool\footnote{http://irsa.ipac.caltech.edu/applications/finderchart}
from NASA's Infrared Science Archive (IRSA). We also incorporated the
diagnostic information provided by the various K2 data reduction
pipelines.  For those targets where source confusion is a concern, we
either took the existing LC most likely to be centered on the target
given the diagnostic information provided by the corresponding
pipelines, or re-reduced the data using the Cody approach, but with a
small, fixed aperture.  

We omitted two highest-quality members (see
Sec.~\ref{sec:membership}): EPICs  247031436 (HD 28867B; confused with
nearby sources) and EPIC 247989931 (2MASS J04412464+2543530; too
faint).  For completeness, we note that EPIC 211068851 (HD 26212) is
too bright and EPIC 210893410 is too faint, but both are also unlikely
to be Taurus or foreground members.

There are a few targets that are either erroneously listed as distinct
targets in the EPIC catalog or could be a faint source next to a
brighter source; in either case, a good LC can't be obtained. HL Tau
(EPIC 210690913) and XZ Tau AB (EPIC 210690892) are high quality
members and close together on the sky, though resolvable; the XZ Tau
binary itself isn't resolved. An additional target is listed as EPIC
210690886, but it is too faint and/or close to XZ Tau to measure
independently. Given its location, position matching with literature
data (see Sec.~\ref{sec:litphotom} below) accidently picks up several
matches that should have been matched to XZ Tau AB. As a result, EPIC
210690886 is explicitly removed from our database as effectively a
duplicate.  DR Tau (EPIC 246923113) is also a high-quality member. 
EPIC 246923117 is 0.22 arcsec away from DR Tau, and is explicitly
removed as effectively a duplicate. EPIC 246926943 is a real though
optically faint target near a brighter source. The LC isn't easily
distinguishable from the bright neighbor. This target is listed as a
NM as a result.

\subsection{Membership} 
\label{sec:membership}

\subsubsection{Summary}

The details of our membership selection process are given  in
Appendix~\ref{app:membership}. To summarize,  we have selected 156
``highest confidence'' Taurus members, 23 ``possible'' Taurus members,
and 38 stars that are an older ($\sim$30 Myr) ``foreground''
population. The highest confidence members are those objects listed as
Taurus members in Esplin \& Luhman (2019). The possible members are
additional objects selected individually by us from the available
literature and Gaia DR2 kinematic data as Taurus members; these are
generally more dispersed than the highest confidence members. The
$\sim$30 Myr foreground population includes (but is not limited to)
members of Group 29 (Oh \etal\ 2017, Luhman 2018); some stars might
potentially belong to Mu Tau (Gagne \etal\ in prep, Liu \etal\ 2020),
or possibly Cas-Tau (e.g., Hartmann  \etal\ 1991, de Zeeuw \etal\
1999, Luhman 2018; also see appendix of David \etal\ 2018 and
references therein). 

Membership status is used in Table~\ref{tab:summarystats} for the
sample breakdown, and is included in Table~\ref{tab:bigperiods}. The
highest confidence sample is listed as distinct from the possible
member sample. The highest confidence and possible member sample is
combined into one sample listed as the $\sim$3 Myr member sample. The
foreground population is a distinct sample with different kinematics
and mean distance, and likely a significantly older age than the
Taurus population. For completeness, Table~\ref{tab:summarystats} also
includes the NM sample. Appendix~\ref{app:nm} lists individually the
information on stars in the probable NM. 

Unless explicitly indicated, the subsequent analysis in this paper
uses the 156 most confident members (typically colored green in
figures) plus the 23 possible members (typically colored yellow-orange
in figures) as the member sample. The 38 stars in the foreground older
population are colored blue in figures.

\subsubsection{Sample Statistics}
\label{sec:samplestatistics}

In general, we expect that Taurus members (in comparison to NM) will
have a higher fraction of disk indicators, have a Gaia DR2 parallax
between about 6 and 8 mas, and have a high fraction of periodic LCs.
In practice, since efforts to determine spectral types have focused on
members, we also expect members to have a higher fraction of spectral
types in the literature. As can be seen in
Table~\ref{tab:summarystats}, all of this is borne out by the Taurus
member sample. There is a high fraction of spectral types
(Sec.~\ref{sec:litphotom}), periodic stars (Sec.~\ref{sec:periods}),
long-wavelength IR detections and therefore disked stars
(Sec.~\ref{sec:finddisks}) among the members. 

The foreground sample, because it is older ($\sim$30 Myr rather than
$\sim$3), has a lower fraction of disks (and IR detections). It also
has a higher fraction of periodic stars, but this is highly biased,
since we started from the set of K2 LCs in C4/C13, not from the
set of all possible members of this foreground population (which may
extend over more K2 campaigns). 

The reddening as determined above (Sec.~\ref{sec:dereddening}) shows
that for the highest confidence members, the distribution of $A_K$
peaks at $\sim$0.1 with a tail to $\sim$2; both the groups of possible
members and the foreground have $ A_K < 0.3$, with a peak $<$0.05. 

Among our K2 work in older clusters, a very high fraction of the
members are periodic -- 92\% in Pleiades (Papers I-II), 87\% in
Praesepe (Paper IV), and 86\% in USco (Paper V). We theorized that the
lower fractions in Praesepe and USco may be indicative of NM
contamination.  In Taurus, 81\% of the highest quality members stars
are periodic; in \rop, only 61\% are periodic. Two things are likely
contributing to the lower fraction of periodic stars in these youngest
clusters. Disk emission affects period determination in that
stochastic contributions from the disk make periods harder to find.
And, NM contamination is likely to also be a factor. Much more effort
has been devoted to determining Taurus membership than that for \rop,
so it is more likely that there is a higher NM contamination rate in
\rop.  The disk fraction among \rop\ member stars with a K2 LC is
46\%, very similar to the  Taurus member sample of $\sim$56\%
(Table~\ref{tab:summarystats}); \rop\ is thought to be younger
($\sim$1 Myr) than Taurus ($\sim$3 Myr), and thus should have more
disks. In both cases, most of the disked stars are also periodic. We
suspect that the NM disk-free contamination is higher in \rop\ than
Taurus, at least partially accounting for the relatively low fraction
of periodic stars in \rop. In contrast, the membership is likely
better in Taurus, and the lower fraction of periodic stars arises
primarily from `pollution' of the LC by disk-related effects, making
periods harder to find. 

We note as well that the lowest mass Taurus members studied here
extend into the brown dwarf regime (see, e.g., Scholz \etal\ 2018).
These mid-Ms represent the youngest, lowest mass stars with high
quality LCs yet observed.

\subsection{Literature Data}
\label{sec:litphotom}

The literature on stars in (or around or near) Taurus is rich indeed.
Aiming for uniform consistency rather than completeness, rather than
scouring the literature for individual photometric measurements, we
assembled information on each target from several all-sky or
large-scale surveys. We matched by position to each catalog with
typical matched positions within an arcsecond. 

The optical data came from several catalogs. We searched in Gaia DR1
(Gaia Collaboration 2016) for the $G$ magnitude consistent with our
prior analyses, and DR2 (Gaia Collaboration 2018) for the more recent
$G$, $R_p$, and $B_p$ photometry, in addition to parallaxes and proper
motions. The kinematic data provided by Gaia DR2 are particularly
useful in identifying members (Sec.\ \ref{sec:membership} includes the
Gaia DR2 ID; see Table~\ref{tab:summarystats} for fractions of samples
with Gaia parallaxes). The distances we used here are those provided
by Bailer-Jones \etal\ (2018).  Additional optical data were obtained
from the AAVSO Photometric All-Sky Survey (APASS; Henden \etal\ 2016),
specifically $V$ magnitudes. We used both Pan-STARRS1 (PS1; Chambers
\etal\ 2016) and the Sloan Digital Sky Survey (SDSS; e.g., Ahn \etal\
2014) for multi-band optical data.  Every star here has several
optical measurements. 

We used infrared data from three all-sky surveys: the Two-Micron All
Sky Survey (2MASS; Skrutskie \etal\ 2006) at $J$, $H$, and $K_{\rm
s}$; the Widefield Infrared Survey Explorer (WISE; Wright \etal\ 2010)
at 3.5, 4.6, 12, and 22 \mum; and AKARI (Murakami \etal\ 2007) data at
9, 18, 65, 90, 140, and 160 \mum.  WISE and AKARI have some similar
wavelengths, but have very different sensitivities.  
We also used infrared data from two pointed
missions (as opposed to all-sky surveys).  For the Spitzer Space
Telescope (Werner \etal\ 2004), we started with data from the Spitzer
Enhanced Imaging Products,
SEIP\footnote{http://irsa.ipac.caltech.edu/data/SPITZER/Enhanced/SEIP/overview.html}
at 3.6, 4.5, 5.8, 8, and 24 \mum, which aggregates all data from the
cryogenic portion of the Spitzer mission. The SEIP only includes
objects with high signal-to-noise ratio detections; when flux
densities did not appear in the SEIP for our targets, we used catalog
data from the Spitzer Taurus project (Rebull \etal\ 2010) for any
bands between 3.6 and 70 \mum\ (note that 70 \mum\ was not included in
the SEIP, so all 70 \mum\ data come from the Spitzer Taurus project).
Most of the members have a Spitzer counterpart; a relatively low
fraction of the NM have a Spitzer counterpart (see
Table~\ref{tab:summarystats}). Relatively few stars in the sample have
a Spitzer/MIPS counterpart at 24 or 70 \mum.  Herschel data cover a
portion of the Taurus region. Analogous to the SEIP, the Herschel
Space Observatory (Pilbratt \etal\ 2010) Highly Processed Data
Products (HPDP) for PACS 70, 100, and 160 \mum\ (Marton \etal\ 2017)
incorporate all the data from the mission.  Very few targets have a
PACS counterpart, but a higher fraction of the members are detected (a
rate that is biased by the observations that targeted known members).
K2 generally monitored optically bright stars, thereby selecting
against embedded stars bright in the mid- and far infrared. 

In general, there is a steep drop off in counterpart numbers as
wavelength increases. Table~\ref{tab:summarystats} includes the
numbers and fractions of stars in several membership categories (Sec.\
\ref{sec:membership}) for several representative infrared bands.
However, the fraction of Taurus member stars (not foreground, not
NM) that are detected out to long wavelengths is much higher
than the fraction of NM stars, as expected -- Taurus member
stars are more likely to have an IR excess, and thus be detected in
these various infrared data sets. 

We used all of these photometric data to assemble a spectral energy
distribution (SED) for each target.  If the data from one catalog were
obviously inconsistent with the rest of the SED, then we removed the
data points from that catalog for that source on the assumption that
the positional match failed. We also used the IRSA Finder Chart tool
(as well as its sister tool, IRSA Viewer) to investigate source
mismatches.

Our earlier papers use \vmkz\ as a proxy for mass, so to better enable
inter-cluster comparisons, we continue to do that here. All the stars
have measured \ks, leaving us to find or calculate $V$. A substantial
fraction of the targets have measured $V$ magnitudes we could assemble
from the literature above, but several stars still lack $V$
measurements. If a Gaia $G$ magnitude is available (from DR1), then
$(V-K_s)$ was interpolated from $(G-K_s)$ as in paper IV (which used
DR1 Gaia data to establish this relationship); we estimate errors on
these estimates to be $\sim$0.017-0.085 mag. For stars redder than
$(V-K_s)\sim5$, the relation from paper IV is linearly extrapolated to
$(V-K_s)\sim8$. Similarly, if no Gaia $G$ mag is available, but a
Pan-STARRS1 $g$ is available, then $(V-K_s)$ can be calibrated via an
empirical relation between $(g-K_s)$ and \vmk; errors on these
estimates are probably comparable to those from Gaia-derived colors.
As a last resort, for those stars still missing a $V$ estimate, since
the SED is well-populated in the optical using literature photometry,
a $V$ magnitude is linearly interpolated from the SED.  All the
targets thus have a measured or inferred \vmk. 

Table~\ref{tab:bigperiods} includes, for the most
likely members, possible members, and foreground population
(identified in section \ref{sec:membership}), the relevant supporting
photometric data, including the observed or interpolated \vmk, plus
the periods we derive (in Section~\ref{sec:periods}) and the IR excess
assessments (Sec.~\ref{sec:finddisks}).  A similar table with all the
likely NM appears in Appendix~\ref{app:nm}.

Most of the Taurus members have spectral types in the literature
(Table~\ref{tab:summarystats}). We assembled spectral types largely
from Luhman (2017, 2018) and Esplin \& Luhman (2017, 2019); some
individual sources have spectral types not collected into these
papers, and we assigned those spectral types individually based on the
literature.

For those stars with spectral types, we used the spectral types as
part of the process for identifying disks (see
Sec.~\ref{sec:finddisks}) and determining the best value of reddening
(see Sec.~\ref{sec:dereddening}). For those stars with spectral types,
we used a Kurucz-Lejeune (Kurucz 1993; Lejeune \etal\ 1997) model
corresponding to that spectral type and normalized to the data at $J$
band. We extended a Rayleigh-Jeans line from the longest wavelength
model data point to past 24 \mum. Using the reddening law from Mathis
(1990), we reddened the models for a grid of $A_J$ values, calculated
a reduced chi-sq ($\chi_{\nu}^2$) comparing the model to the data at
all bands $J$ and shorter wavelength, and selected the best fit based
on the smallest $\chi_{\nu}^2$. These model fits to the SED are not
meant to be rigorous, but instead `guide the eye.'

\floattable
\begin{deluxetable}{rcccccc}
\tabletypesize{\scriptsize}
\tablecaption{Summary of statistics on Taurus sample\tablenotemark{a}\label{tab:summarystats}}
\tablewidth{0pt}
\tablehead{\colhead{property} & \colhead{high-confidence}
&\colhead{possible}& \colhead{$\sim$3 Myr\tablenotemark{b}}& 
\colhead{ $\sim$30 Myr \tablenotemark{c}}& \colhead{ NM or
rejected}& \colhead{ all} \\ [-0.4cm]
\colhead{} & \colhead{members} & \colhead{members}  & \colhead{members}&
 \colhead{foreground} & \colhead{sources}& \colhead{sources}}
\startdata
C4 &            4 &           3 &           7 &          23 &         131 &         161\\
C13 &          152 &          22 &         174 &          16 &         513 &         703\\
C4 AND C13 &            0 &           2 &           2 &           1 &          10 &          13\\
C4 OR C13 &          156 &          23 &         179 &          38 &         634 &         851\\
\hline
all &          156 &          23 &         179 &          38 &         634 \\
periodic    &          127 (0.81) &          21 (0.91) &         148 (0.83) &          37 (0.97) &         307 (0.48) \\
multi-periodic &           32 (0.21) &           8 (0.35) &          40 (0.22) &          12 (0.32) &          81 (0.13) \\
no IR excess &           55 (0.35) &          16 (0.70) &          71 (0.40) &          36 (0.95) &         585 (0.92) \\
high-conf IR excess &           94 (0.60) &           7 (0.30) &         101 (0.56) &           1 (0.03) &          35 (0.06) \\
possible IR excess &            7 (0.04) &           0 (0.00) &           7 (0.04) &           1 (0.03) &          14 (0.02) \\
IR excess+periodic &           67 (0.43) &           5 (0.22) &          72 (0.40) &           1 (0.03) &           8 (0.01) \\
IR excess+multi-periodic &           11 (0.07) &           2 (0.09) &          13 (0.07) &           0 (0.00) &           2 (0.00) \\
burster &           22 (0.14) &           3 (0.13) &          25 (0.14) &           0 (0.00) &           0 (0.00) \\
burster+IR excess &           22 (0.14) &           3 (0.13) &          25 (0.14) &           0 (0.00) &           0 (0.00) \\
burster+IR excess+periodic &           14 (0.09) &           2 (0.09) &          16 (0.09) &           0 (0.00) &           0 (0.00) \\
dipper &           18 (0.12) &           3 (0.13) &          21 (0.12) &           0 (0.00) &           0 (0.00) \\
dipper+IR excess &           18 (0.12) &           3 (0.13) &          21 (0.12) &           0 (0.00) &           0 (0.00) \\
dipper+IR excess+periodic &           16 (0.10) &           2 (0.09) &          18 (0.10) &           0 (0.00) &           0 (0.00) \\
\hline
spectral type &          151 (0.97) &          22 (0.96) &         173 (0.97) &          26 (0.68) &         155 (0.24) \\
Gaia parallax &          129 (0.83) &          20 (0.87) &         149 (0.83) &          38 (1.00) &         558 (0.88) \\
IRAC-1    &          135 (0.87) &           4 (0.17) &         139 (0.78) &           6 (0.16) &         168 (0.26) \\
WISE-1    &          156 (1.00) &          23 (1.00) &         179 (1.00) &          38 (1.00) &         633 (1.00) \\
AKARI 18 &           39 (0.25) &           2 (0.09) &          41 (0.23) &           0 (0.00) &          14 (0.02) \\
MIPS-24       &           98 (0.63) &           9 (0.39) &         107 (0.60) &          12 (0.32) &          85 (0.13) \\
MIPS-70   &           40 (0.26) &           2 (0.09) &          42 (0.23) &           0 (0.00) &           4 (0.01) \\
PACS 70  &           64 (0.41) &           2 (0.09) &          66 (0.37) &           0 (0.00) &           4 (0.01) \\
\hline
$V$ and \ks\ measured        &          57 (0.37) &          17 (0.74) &          74 (0.41) &          22 (0.58) &         207 (0.33) \\
$V$ from APASS               &          19 (0.12) &           4 (0.17) &          23 (0.13) &           8 (0.21) &         262 (0.41) \\
\vmk\ via Gaia $G -$\ks      &          36 (0.23) &           2 (0.09) &          38 (0.21) &           5 (0.13) &         145 (0.23) \\
\vmk\ via PanSTARRS $g-$\ks  &          34 (0.22) &           0 (0.00) &          34 (0.19) &           2 (0.05) &          13 (0.02) \\
SED-interpolated $V$         &          10 (0.06) &           0 (0.00) &          10 (0.06) &           1 (0.03) &           7 (0.01) \\
\hline
$A_v$ from $JH$\ks\ diagram  &           31 (0.20) &           4 (0.17) &          35 (0.20) &           9 (0.24) &         313 (0.49) \\
$A_v$ from spectral type     &          115 (0.74) &          15 (0.65) &         130 (0.73) &          16 (0.42) &         107 (0.17) \\
$A_v$ from SED fits          &            5 (0.03) &           4 (0.17) &           9 (0.05) &          11 (0.29) &          25 (0.04) \\
median $A_v$ assigned        &            5 (0.03) &           0 (0.00) &           5 (0.03) &           2 (0.05) &         189 (0.30) \\
\enddata
\tablenotetext{a}{Numbers in table are raw number of stars meeting
the stated criterion/criteria, followed by the sample fraction in
parentheses.}
\tablenotetext{b}{$\sim$3 Myr members are the high-confidence (Taurus) member
plus the possible (Taurus) member sample.}
\tablenotetext{c}{The $\sim$30 Myr foreground sample are
stars determined in the literature to be young, but not kinematically
consistent with Taurus membership; these could be members of, e.g.,
Group 29 (see Sec.~\ref{sec:membership}).}
\end{deluxetable}

\clearpage

\floattable
\begin{deluxetable}{ccp{13cm}}
\tabletypesize{\scriptsize}
\tablecaption{Contents of Table: Periods and Supporting Data for
Taurus Members with K2 Light Curves\label{tab:bigperiods}}
\tablewidth{0pt}
\tablehead{\colhead{Number} & \colhead{Column} & \colhead{Contents}}
\startdata
1 & EPIC & Number in the Ecliptic Plane Input Catalog (EPIC) for K2\\
2 & coord & Right ascension and declination (J2000) for target \\
3 & othername & Alternate name for target \\
4 & gaiaid & Gaia DR2 ID \\
5 & Vmag & V magnitude (in Vega mags), if observed\\
6 & Kmag & \ks\ magnitude (in Vega mags), if observed\\
7 & vmk-obs & \vmk, as directly observed (in Vega mags), if $V$ and \ks\ exist\\
8 & vmk-used & \vmk\ used, in Vega mags (observed or inferred; see text)\\
9 & evmk & $E(V-K_s)$ adopted for this star (in mags; see \S~\ref{sec:dereddening}) \\
10 & Kmag0 & dereddened $K_{s,0}$ magnitude (in Vega mags), as inferred (see \S\ref{sec:dereddening})\\
11 & vmk0 & $(V-K_s)_0$, dereddened $V-K_s$ (in Vega mags), as inferred (see \S~\ref{sec:dereddening}; rounded to nearest 0.1 to emphasize the relatively low accuracy)\\
12 & uncertaintycode & two digit code denoting origin of \vmk\ and \vmkz\
(see \S\ref{sec:litphotom} and \ref{sec:dereddening}):
First digit (origin of \vmk): 
1=$V$ measured directly from the literature (including SIMBAD) and $K_s$ from 2MASS; 
2=$V$ from APASS and $K_s$ from 2MASS;
3=\vmk\ inferred from Gaia $g$ and $K_s$ from 2MASS (see \S\ref{sec:litphotom});
4=\vmk\ inferred from Pan-STARRS1 $g$ and $K_s$ from 2MASS (see \S\ref{sec:litphotom});
5=\vmk\ inferred from membership work (see \S\ref{sec:membership}; rare); 
6=$V$ inferred from well-populated optical SED and $K_s$ from 2MASS (see \S\ref{sec:litphotom});
-9= no measure of \vmk.
Second digit (origin of $E(V-K_s)$ leading to \vmkz): 
1=dereddening from $JHK_s$ diagram (see \S\ref{sec:dereddening});
2=dereddening back to \vmkz\ expected for spectral type;
3=used median $E(V-K_s)$=0.7 (see \S\ref{sec:dereddening});
-9= no measure of  $E(V-K_s)$ \\
13 & P1 & Primary period, in days (taken to be rotation period in cases where there is $>$ 1 period)\\
14 & P2 & Secondary period, in days\\
15 & P3 & Tertiary period, in days\\
16 & P4 & Quaternary period, in days\\
17 & Membership & Highest quality member, possible member, foreground young star (see \S\ref{sec:membership}) \\
18 & IRexcess & Whether an IR excess is present or not (see \S\ref{sec:disks})\\
19 & IRexcessStart & Minimum wavelength at which the IR excess is detected or the limit of our knowledge
of where there is no excess (see \S\ref{sec:disks}) \\
20 & slope & Slope of SED fit to all available detections between 2 and 25 \mum\ \\
21 & SEDclass & SED class (I, flat, II, III) \\
22 & dipper & LC matches dipper characteristics (see \S\ref{sec:LCPcats})\\
23 & burster & LC matches burster characteristics (see \S\ref{sec:LCPcats})\\
24 & single/multi-P &  single or multi-period star \\
25 & dd &  LC and power spectrum matches double-dip characteristics (see \S\ref{sec:LCPcats})\\
26 & ddmoving & LC and power spectrum matches moving double-dip characteristics (see \S\ref{sec:LCPcats})\\
27 & shapechanger & LC matches shape changer characteristics (see \S\ref{sec:LCPcats})\\
28 & beater &  LC has beating visible (see \S\ref{sec:LCPcats})\\
29 & complexpeak & power spectrum has a complex, structured peak and/or has a wide peak (see \S\ref{sec:LCPcats})\\
30 & resolvedclose & power spectrum has resolved close peaks (see \S\ref{sec:LCPcats})\\
31 & resolveddist & power spectrum has resolved distant peaks (see \S\ref{sec:LCPcats})\\
32 & pulsator & power spectrum and LC match pulsator characteristics (see \S\ref{sec:LCPcats})\\
\enddata
\end{deluxetable}

\subsection{Disk Indicators}
\label{sec:finddisks}

A substantial fraction of the Taurus members have an IR excess, from
which we can infer the presence of a circumstellar disk (see summary
statistics in Table~\ref{tab:summarystats}). In the context of this
paper (as for Paper V), we wish to have a {\em complete} list of
disks, as opposed to an unbiased list; we identify a star as a disk
candidate if it has a plausibly real excess at any IR wavelength (at
which we have a detection) in the catalog we assembled.  The
wavelength at which the IR excess begins is included in
Table~\ref{tab:bigperiods} (and in the Appendix in
Table~\ref{tab:bignm} for the NM).

Because WISE is lower spatial resolution than Spitzer, we used IRSA's
Finder Chart tool to inspect the WISE images to see if the detections
in the catalog reflect what can be seen in the images. Again following
Paper V, to identify disks, we looked at the significance of any
putative IR excess at 12, 22, and 24 \mum\ where available (with
$\chi$ calculated as described in paper V), taking into account that
for the latest types, the expected photospheric colors may be $>$0. We
used the empirical photospheric infrared color [W1]$-$[W3] as a
function of \vmkz\ from Paper V.  We then assessed the ensemble of
information available for all sources (e.g., all points $\geq$2 \mum,
the shape of SED, the results of the simple model fit discussed in
Sec.~\ref{sec:litphotom} above, etc.). For each source, we have an
assessment of whether it has a disk, and if so, the shortest
wavelength likely contributing to an IR excess. 

In this fashion, as for Paper V, we identified unambiguous disk
candidates and non-disk candidates (at least, non-disks given the
available data, which often extend at least to 12 or 22 \mum), with a
few percent of ambiguous (possible) disks noted as such.
Table~\ref{tab:summarystats} includes number and sample fraction in
each of the relevant samples. More than half of the Taurus members
have disks; there is just one high confidence (and one more possible)
disk candidate among the foreground ($\sim$30 Myr) population. Note
that the disk excess criteria are conservative and that the non-disked
sample will likely have contamination from weaker ($<5-10\sigma$
excess) disks. Note also that the lowest mass bin is likely incomplete
in the non-disks due to sensitivity issues (stars with excesses are
more likely to be detected at long wavelengths than stars without
excesses). While this sample draws from many surveys and wavelengths,
in order to be considered at all, there must be an observation in K2,
which requires targeting of the source by a human, and therefore the
sample is affected not only by extinction but pixel mask selection.

Independent of the disk candidate status, we performed a simple
ordinary least squares linear fit to all available photometry (with
errors, but just detections, not including upper or lower limits)
between 2 and 24 $\mu$m, inclusive.  In the spirit of Wilking \etal\
(2001), we define $\alpha = d \log \lambda F_{\lambda}/d \log 
\lambda$, where  $\alpha > 0.3$ for a Class I, 0.3 to $-$0.3 for a
flat-spectrum  source, $-$0.3 to $-$1.6 for a Class II, and $<-$1.6
for a Class III.  The slope and the class are both available in
Table~\ref{tab:bigperiods} (and Table~\ref{tab:bignm}, where it is
assumed that the IR excess is due to a circumstellar disk, which may
not be a good assumption).

For completeness, we note two items. First,
Table~\ref{tab:summarystats} indicates that there are NM with IR
excesses. These objects include evolved stars with dusty winds or
envelopes, distant Be stars, and a few extragalactic objects (AGN,
QSO). Second, Table~\ref{tab:bigperiods} (and Table~\ref{tab:bignm})
indicates the onset of IR excess for each target. Since the IR excess
could affect the \ks\ bands, which could cause an overestimation of
\av\ in the methods described below, we note that such targets are a
minority here; $>$80\% of the highest quality members have disks that
start at longer than 3 \mum.

\subsection{Dereddening}
\label{sec:dereddening}

The reddening in the direction of Taurus can be substantial, and
patchy. As noted above, we obtained spectral types from the
literature.  In order to deredden the $V-K_s$ colors, we followed the
same approach as in paper V (specifically for $\rho$ Oph), with some
modifications.

We can place essentially all of the stars on a $J-H$ vs.\ $H-K_s$
diagram. We can deproject much of the sample back along the reddening
law derived by Indebetouw \etal\ (2008) to the expected $JHK_s$ colors
for young stars from Pecaut \& Mamajek (2013) or the T~Tauri locus
from Meyer \etal\ (1997). Note that there is a discontinuity between
the end of the Pecaut \& Mamajek relation and the beginning of the
T~Tauri locus (noted in Meyer \etal\ 1997 and paper V); this results
in a small gap in the dereddened $(J-K_s)_0$ distibution between
$\sim$0.9 and $\sim$1.0. The reddening so derived can be converted to
$E(V-K_s)$ via $A_K = 0.114 A_V$ (Cardelli \etal\ 1989). This approach
worked well for Upper Sco in paper V, but did not work as well in
$\rho$ Oph, primarily because there is significantly higher reddening.
Similarly, there is high reddening towards some of the stars in
Taurus, and there is a lot of scatter, even in the NIR colors.

Another approach used in paper V (and preferred in $\rho$ Oph) was to
take the spectral type, compare the observed colors to those expected
for that type from Pecaut \& Mamajek (2013), and calculate the
resultant reddening. Consequently, any distribution of colors for
stars dereddened in this fashion will be quantized,
because, e.g., all stars of type M1 are assigned to have the same color.
Due to the high fraction of spectral types for Taurus
members, we can calculate reddening following the above prescription
for most of the members.  

Because there is patchy reddening and a lot of scatter in the $JH$\ks\
diagram (much more than for USco, for example), we took the reddening
estimate from the spectral type first. If that value was unphysical
(e.g., $<$0), then we took the value derived from the  $JH$\ks\
diagram. Some of those values were still unphysical or insufficient
(e.g., \vmkz\ was still $>$ 8 or 10). So, for those stars with
spectral types, we used the Kurucz-Lejeune model grid fitting
described above (Sec.~\ref{sec:litphotom}) to estimate the reddening
in a third way. For several of the foreground population lacking
spectral types, the reddening derived from the $JH$\ks\ diagram was
substantially discrepant from the SED fitting having guessed a
spectral type based on the observed \vmk; in those cases, we selected
the best reddening from the SED fitting. As a fourth and last resort,
we assigned a star the modal reddening of $E(V-K_s)\sim$0.46
determined via the comparison of observed values to that expected from
the spectral type for the ensemble.

Table~\ref{tab:summarystats} includes the numbers and sample fractions
for each of these approaches. The value used for most stars comes from
either of the first two approaches (spectral type or $JH$\ks\
diagram). For relatively few stars do we have to fall back to the most likely
reddening.    

The dereddened \vmkz\ we used for each object is included in
Table~\ref{tab:bigperiods} for the members and in
Appendix~\ref{app:nm} for the NM. However, to emphasize the net 
uncertainty, the ``vmk0'' column in Table~\ref{tab:bigperiods} 
has been rounded to the nearest 0.1 mag. The values used in plots 
here can be recovered by using the $E(V-K_s)$ (``ev-k'') and 
$(V-K_s)_{\rm observed}$ (``vk-used'') columns. 

Net errors are hard to quantify after all of these steps.
Table~\ref{tab:bigperiods} (and its analogous Table~\ref{tab:bignm}
for NM) include a 2-digit code indicating the origin of the \vmk\
value and the method by which the \vmk\ was dereddened to \vmkz\ (see
Table~\ref{tab:bigperiods} or \ref{tab:bignm} for specific
definitions). The reddening can be large and a significant source of
uncertainty.  Via internal comparisons and uncertainties not just on
the assumed photospheric colors but also uncertainties in spectral
typing, we estimate that the typical uncertainty for Taurus members
could be as much as 1 magnitude in $E(V-K_s)$.  To quantify this
further, for the highest  quality members, we compared the reddening
values derived from the $JH$\ks\ diagram, from spectral types, and
from SED fitting. The reddening derived from the SED
is well-matched to that from the $JH$\ks\ diagram; that from the
spectral type has systematic offsets with respect to the others in the
sense that the reddening from the spectral types is lower on average,
but also negative (e.g., unphysical) far more often. In all cases,
a Gaussian fit to the distribution of fractional differences 
suggests a scatter of $\sim$0.5 mag in $E(V-K_s)$ is typical. Most of the $JH$\ks\
measurements come from 2MASS, where the measurements should have been
roughly simultaneous for the three bands, so intrinsic variability
alone cannot account for the scatter. It is not the case that emission
from the disk is affecting the $JH$\ks, because the K2 sample is
biased towards optical sources, and as noted above, most ($>$80\%) of
these disks become apparent at 3.5 \mum\ or longer. The large
uncertainty in reddening is real but evidently unavoidable.

\section{Periods and Color-Magnitude Diagrams}
\label{sec:interp}

\subsection{Finding Periods in the K2 LCs}
\label{sec:periods}

Our approach for finding periods was identical to that we used
in Papers I, II, IV, and V. In
summary, we used the Lomb-Scargle (LS; Scargle 1982) approach as
implemented by the NASA Exoplanet Archive Periodogram
Service\footnote{http://exoplanetarchive.ipac.caltech.edu/cgi-bin/Periodogram/nph-simpleupload}
(Akeson \etal\ 2013). We also used the Infrared Science Archive (IRSA)
Time Series
Tool\footnote{http://irsa.ipac.caltech.edu/irsaviewer/timeseries},
which employs the same underlying code as the Exoplanet Archive service,
but allows for interactive period selection. We looked for periods
between 0.05 and 35 d, with the upper limit being set by roughly half
the campaign length. Because the periods are typically unambigous,
false alarm probability (FAP) levels are calculated as exactly 0 for
most of the periods we present here (and the remaining FAP levels are
typically $<10^{-4}$).  

The periods we derive appear in Table~\ref{tab:bigperiods} 
for all members and in Appendix~\ref{app:nm} for the NM.

\begin{figure}[ht]
\epsscale{1.0}
\plotone{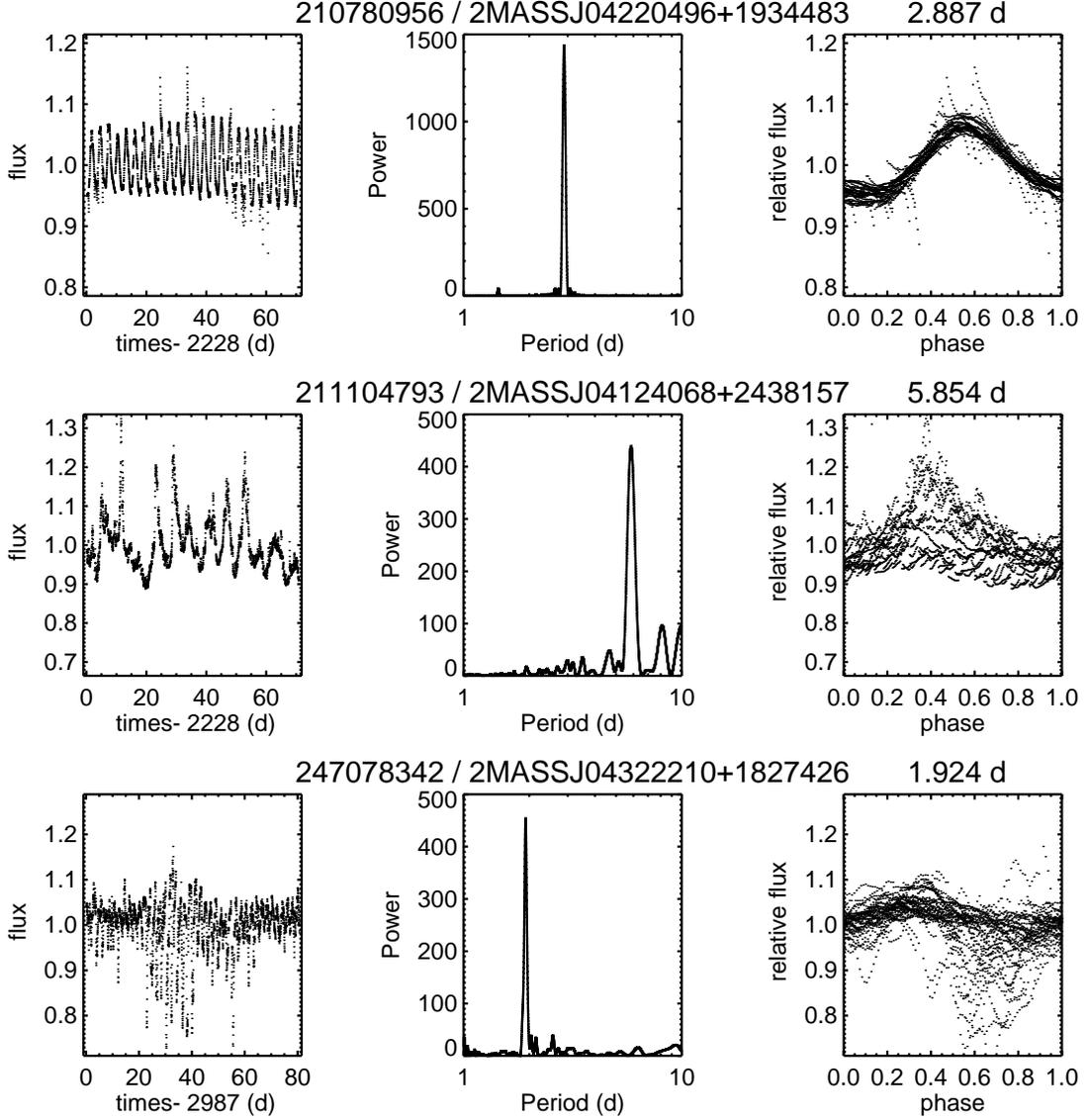}
\caption{Example Taurus LCs for a single-period sinusoid LC (first row),
burster (second row), and dipper (third row). The panels are: best LC,
power spectrum, phased LC for first period.  The objects are, top to
bottom, EPIC 210780956/2MASSJ04220496+1934483 (no IR excess; data from
C4), EPIC 211104793/2MASSJ04124068+2438157 (large IR excess; data from
C4), and EPIC 247078342/2MASSJ04322210+1827426 (large IR excess; data
from C13). All of these targets are confident members of Taurus.}
\label{fig:lcexamples}
\end{figure}

\subsection{Interpretation of Periods}
\label{sec:interpofperiods}

In Papers I, II, and IV, we described in detail the LC and periodogram
shapes we observed in the older Pleiades and Praesepe clusters; Paper
V included an additional section on the physical interpretation of the
LCs of younger stars in USco and $\rho$ Oph. Here we briefly
summarize, and refer the reader to those earlier papers for more
detail.

Including both C4 and C13, 81\% (127 stars) of the highest-confidence
Taurus member sample stars have at least one identifiable period in
their LC. We retain up to four periods; 32 (21\%) have more than one
period. Nearly all of the foreground population has at least one
period. More than 95\% of the sample we focus on here (members of
Taurus or belonging to the foreground) have LCs consistent with a
surface spot origin.   For stars with two (or more) periods that we
believe are due to rotation, we plot only one point at the period we
believe corresponds to the actual rotation period of the star
dominating the \vmkz\ measurement.  In those cases, particularly among
the M stars, where there are two (or more) periods that are not close
together, we believe the star to be a likely binary (or other
multiple); see Papers II and IV, and Stauffer \etal\ (2018b). 

In a few stars, just two of the possible members, there is a forest of
very short period peaks in the periodogram. We take these to be
pulsators.

There are 7 stars in the entire sample that have little or no IR
excess, but have periodic LC shapes with scallop shell or flux dip
morphologies, which cannot be due to spots or pulsation; see Stauffer
\etal\ (2017, 2018a) for a discussion of these types of light curve
morphologies. Of these stars, three of them are highest quality
members (EPIC 246938594, EPIC 246969828 = Cl*Melotte25LH19, and EPIC
247794636 = 2MASSJ04321786+2422149), three are possible members (EPIC
246676629 = UCAC4522-000989, 246682490 = UCAC4522-009859, 247343526 =
2MASSJ04405340+2055471), and one is part of the foreground population
(EPIC 246776923).  Because the primary period in these cases is likely
to either be a rotation period or strongly related to a rotation
period, these were retained as rotation periods.

As in other young clusters, there are LCs with dipper or burster
morphologies (see, Figure~\ref{fig:lcexamples} and, e.g., Paper V;
Cody \etal\ 2014; Cody \& Hillenbrand 2018; Stauffer \etal\ 2014,
2015, 2016a). All the LCs we identified as dippers (21 members) or
bursters (25 members) also have disks, and all are members of some
sort (Table~\ref{tab:summarystats}).  These are identified in the
corresponding tables (Table~\ref{tab:bigperiods} and
Appendix~\ref{app:nm}). Note that the period we report as the rotation
period is often but not always also the period of the dips/bursts;
sometimes the dips/bursts align with the sinusoidal modulation, and
sometimes they do not. Roggero \etal\ (2020 in prep) will explore the
dipper population in Taurus.

We categorize `timescales' for LCs that seem to have a repeating
pattern but the pattern does not seem to be due to starspots or other
rotation-related phenomena and therefore are not taken as periodic;
see Appendix~\ref{app:timescales}.

\subsection{Comparison to Literature Periods}
\label{sec:lit}

Detailed comparison between periods in the literature (consisting of a
few to tens of stars per study) and periods we
derive from K2 data is in Appendix~\ref{app:lit}. 

In summary, we conclude that we are recovering most of the periods
reported in the literature. The results are a mix of excellent
agreement, likely harmonics reported in the literature, and for a few,
significant disagreement with the periods derived from the
high-quality K2 data. 

The study with which we have the most disagreement is the KELT
analysis (Rodriguez \etal\ 2017b), where 19/26 do not match. Our
analysis of other clusters in common with KELT produced a much better
match in periods. Perhaps this is not surprising, in particular
for those stars still actively accreting or where the distribution of
spots may be changing on relatively short timescales; changes in the
LC shape of an accreting object can reasonably be expected to change
on $\sim$year timescales even though the stellar rotation rate itself
does not change on short timescales. 

In any event, for each mismatched period, we investigated the prior
period and believe that the period(s) we report is/are the correct
periods for these stars during the K2 campaign(s).

\subsection{Comparison to Literature $v \sin i$}
\label{sec:vsini}

As discussed above in Sec.~\ref{sec:membership}, it is likely that the
relatively low fraction of Taurus members with periods in the K2 data
is due to disk contributions to the LCs, making periods difficult to
extract.  To determine whether or not this results in a bias in our
derived period ($P$) distribution, we compared our periods with the
projected rotational velocities (\vsini) from the literature. We
started with the \vsini\ compilations in Rebull \etal\ (2004) and
G\"udel \etal\ (2007), and then added those from Nguyen \etal\
(2012). 

\begin{figure}[h]
\epsscale{1.0}
\plotone{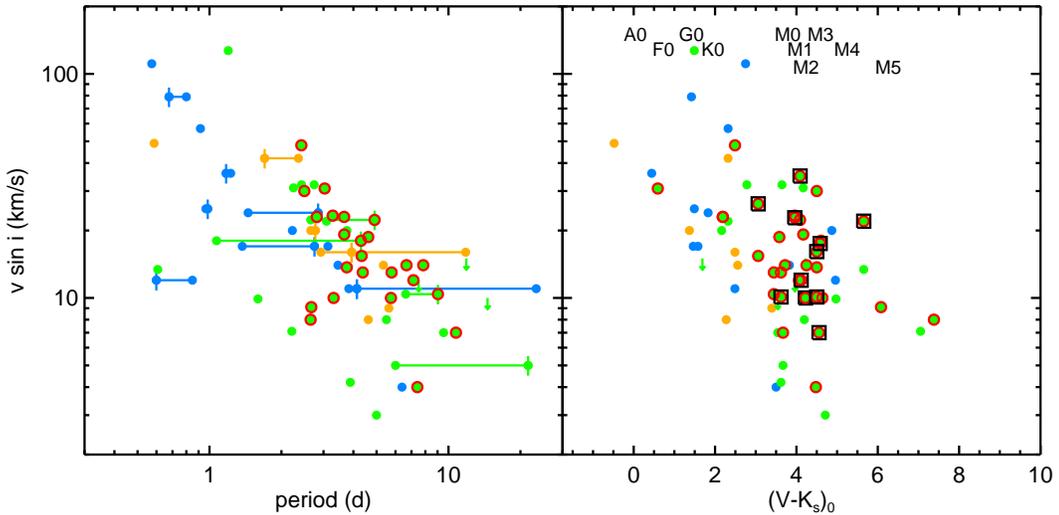}
\caption{Comparison of periods obtained here to \vsini\ from the
literature. Left: \vsini\ (from the literature) vs.\ period (as
derived here). Right: \vsini\ (from the literature) vs. \vmkz. Colors
of points are: green= highest confidence member; orange= possible
member; blue=young, foreground population (see
Sec.~\ref{sec:membership}), where arrows denote upper limits. An
additional red circle around the point denotes a high-confidence disk.
On the left, the points linked by lines connect the first period (what
we have taken as the rotation rate of the star, denoted by a small
vertical line) with the  fastest and slowest period available for that
star, on the assumption that the \vsini\ may not correspond to the
first period for those stars with multiple periods. On the right, an
additional square around the point on the right panel indicates that
we found no period for this star in the K2 data. There is no obvious
trend for the stars lacking periods to have systematically higher or
lower \vsini\ than those with periods.  }
\label{fig:vsini}
\end{figure}

Out of the Taurus members or the foreground population with K2 LCs,
there are 51 with \vsini\ measurements or limits in the literature.
Figure~\ref{fig:vsini} compares the \vsini\ and the periods measured
here. For stars with multiple measured periods, the measured \vsini\
in the literature may not correspond to the period we have taken as
the rotation period. Thus, those stars with multiple periods are
indicated in the plot.  Disks and stars lacking periods in the K2 data
are indicated. 

On the left of Figure~\ref{fig:vsini}, we recover the expected
relationship in that stars with fast $P$ have large \vsini. Stars with
disks lack shorter periods ($<$2 d; see Sec.~\ref{sec:disks} below). 
On the right of Figure~\ref{fig:vsini}, there is no obvious trend for
the stars lacking periods to have systematically higher or lower
\vsini\ than those with periods. We conclude that our sample with
periods is not significantly biased by omitting the $\sim$20\% of
(member) stars (nearly all of which have disks) that do not have
periods, and comparisons to other, older clusters should be
straightforward.

\subsection{Color-Magnitude Diagrams}
\label{sec:cmd}

Figures~\ref{fig:optcmd1}, \ref{fig:optcmd2}, and \ref{fig:optcmd3}
are various versions of the \ks/\vmk\ color-magnitude diagram (CMD) for
the entire sample with each highlighting the stars of interest: the
most likely Taurus members, the possible Taurus members, and the older
foreground population.  Figure~\ref{fig:optcmd1} is the observed CMD,
and Figures~\ref{fig:optcmd2}-\ref{fig:optcmd3} are dereddened
CMDs. Figs~\ref{fig:optcmd1} and \ref{fig:optcmd2} include apparent
and absolute \ks, but not every star has a Gaia distance (see
statistics in Table~\ref{tab:summarystats}), so Fig.~\ref{fig:optcmd3}
is just apparent magnitudes. Fig.~\ref{fig:optcmd3} also distinguishes
the periodic and the disked stars. 
By inspection of these CMDs, we primarily see that reddening is
important, and the foreground population is older than the Taurus
members.

\subsubsection{Taurus Members}

The observed CMD (Fig.~\ref{fig:optcmd1}) shows a very large scatter
for the Taurus members (both high confidence and possible), most
obviously in \vmk\ color. Correcting for distance alone does not
substantially reduce that scatter. (Correcting for distance makes
large changes to the scatter of the NM, and it becomes apparent that
many NM are giants.)  When dereddened (Fig.~\ref{fig:optcmd2}), the
Taurus members (both high confidence and possible) form a more
recognizable cluster sequence, though still with scatter. Some of this
scatter arises from uncertainty in the reddening correction, but some
is likely intrinsic to the Taurus population. Although we are using
\ks, which can be affected by circumstellar disks, because this sample
is defined by, and thus biased towards, targets with K2 LCs, those
stars with disks substantial enough to significantly affect \ks\ are
generally also those that are so embedded as to not have a K2 LC.
Table~\ref{tab:bigperiods} (and its NM counterpart
Table~\ref{tab:bignm}) includes an indication of where the IR excess
starts, if an IR excess is detected. Less than 20\% of the highest
quality member are likely to have their \ks\ value affected by disk
excesses.

The relative lack of highest confidence members towards the higher
masses (\vmkz$\lesssim$4) noted in Section~\ref{sec:membership} is
apparent; many of the possible members (and for that matter foreground
population; see below) do fall in this color range. 

Fig.~\ref{fig:optcmd3} highlights the periodic and disked samples in
the dereddened CMD. (See statistics in Table~\ref{tab:summarystats}.)
Because a high fraction of the members (high confidence and possible)
are periodic, most of the Taurus stars appearing in the dereddened CMD
also appear in the dereddened CMD of stars with periods. Because a
lower fraction of the NM have periods, many NM do not appear in the
CMD of stars with periods; in general, if a target has a period, it is
more likely to be young. Similarly, a high fraction of the disked
population are also members (most often members of Taurus, as opposed
to the foreground population).

\subsubsection{Foreground Population}

\begin{figure}[ht]
\epsscale{1.1}
\plotone{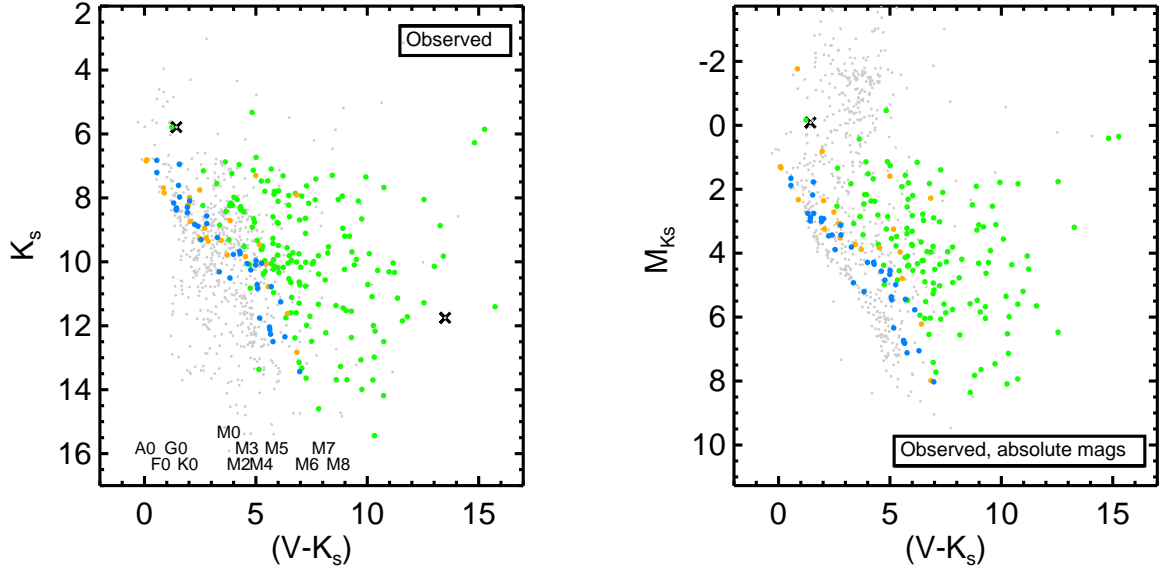}
\caption{Color-magnitude diagrams (\ks\ vs.\vmk) for the sample for
which we had or could infer \vmk\ as described in the text
(\S\ref{sec:litphotom}).  In both plots: green: highest quality
members; orange: possible members; blue: foreground population; grey:
those in the initial sample of candidate members (see
\S\ref{sec:membership}).  Points shown as
$\times$ are too bright or too faint to yield periods from their LCs.
Approximate spectral types for each \vmk\ color are given in the first
plot. Left: Observed values. Note that Taurus members (highest
quality) are often highly reddened, and the foreground population has
a considerably lower range of observed colors.  Right: Observed
values, but shifted for each object for which we had a Gaia DR2
distance (Sec.~\ref{sec:litphotom}, sample fractions in
Table~\ref{tab:summarystats}) to convert the $y$-axis to absolute \ks.
In these observed CMDs, the members have a large dispersion towards
the red. The foreground population is relatively well-behaved. The
NM background includes a lot of giants, as revealed in the
right CMD using absolute magnitudes.}
\label{fig:optcmd1}
\end{figure}

\begin{figure}[ht]
\epsscale{1.1}
\plotone{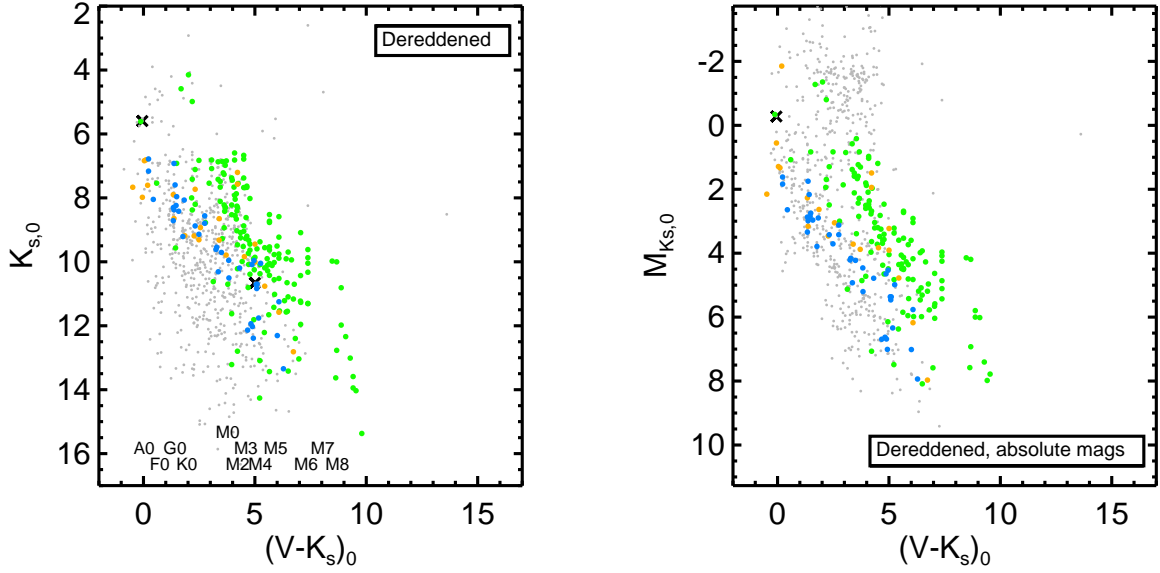}
\caption{Dereddened color-magnitude diagrams (\ks$_{,0}$ vs.\vmkz) for
the sample; notation is as in prior Fig.~\ref{fig:optcmd1}, and note
that the range on the $x$-axis is the same as Fig.~\ref{fig:optcmd1}
as well.  Left: dereddened (\S\ref{sec:dereddening}), observed values;
right:  dereddened absolute values. Correcting for reddening brings
most of the stars bluer, and the change is large for many of the
Taurus member stars in particular. The Taurus members are in a cluster
sequence, but it is not as tight a relationship as in the older
clusters (Papers I or IV) because the stars are so young.  Some
quantization can be seen as a result of the dereddening approach for
some stars with spectral types (see Sec.~\ref{sec:dereddening}).
Little change occurs for the foreground population, as they are less
subject to reddening than the Taurus members. They are lower in the
CMD than the Taurus population, indicating that they are older. }
\label{fig:optcmd2}
\end{figure}

In contrast to the Taurus members in Fig~\ref{fig:optcmd1}, there is
not very much scatter in the foreground population's observed CMD;
these stars on average have low extinction. The foreground sample is
in front of most of the gas and dust associated with Taurus, so this
tight cluster sequence is not surprising. In both
Figs.~\ref{fig:optcmd1} and \ref{fig:optcmd2}, the foreground
population is also clearly lower in the CMD than most of the Taurus members
(from either confidence level), indicating that this foreground
population is older than that of Taurus.

\begin{figure}[ht]
\epsscale{1.2}
\plotone{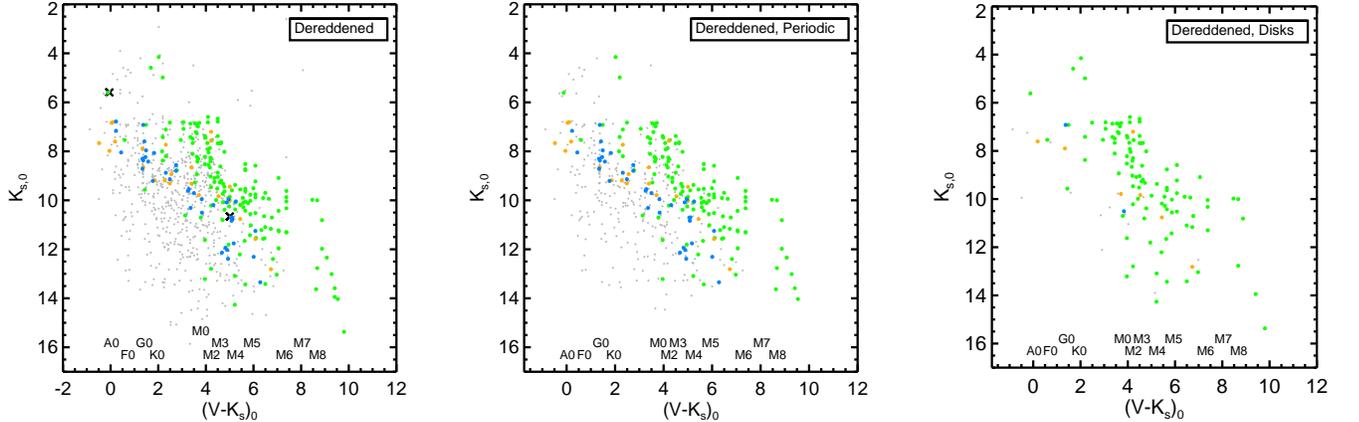}
\caption{Dereddened color-magnitude diagrams (\ks$_{,0}$~vs.\ \vmkz)
for the sample; notation is as in  Fig.~\ref{fig:optcmd1} or
\ref{fig:optcmd2}, but now the $x$-axis is adjusted to encompass the
smaller range of dereddened member \vmkz. Left: \ks$_{,0}$ vs.\ \vmkz\ for the
sample (same as left panel Fig.~\ref{fig:optcmd2}; repeated here to
show with the same $y$-axis as the other plots in this figure).
Middle: \ks$_{,0}$ vs.\ \vmkz\  for the periodic subsample. Right:
\ks$_{,0}$ vs.\ \vmkz\ for the subsample with circumstellar disks (IR
excess). The periodic sample and the sample with disks are more
dominated by Taurus members; nearly all of the foreground population have
periods, and just two have disks.}
\label{fig:optcmd3}
\end{figure}

In Fig.~\ref{fig:optcmd3}, the periodic sub-sample retains all but one
of the stars considered part of the foreground population. The fact
that they have measurable periods means they likely have big spots and
thus are likely young, but they are not Taurus members (see
Sec.~\ref{sec:membership}). Just two of the foreground population have
a disk (one possible disk, one unambiguous disk); this low disk
fraction is another indication that the foreground population is older
than the Taurus members.

\clearpage

\section{Disks and Rotation}
\label{sec:disks}

\begin{figure}[ht]
\epsscale{1.0}
\plottwo{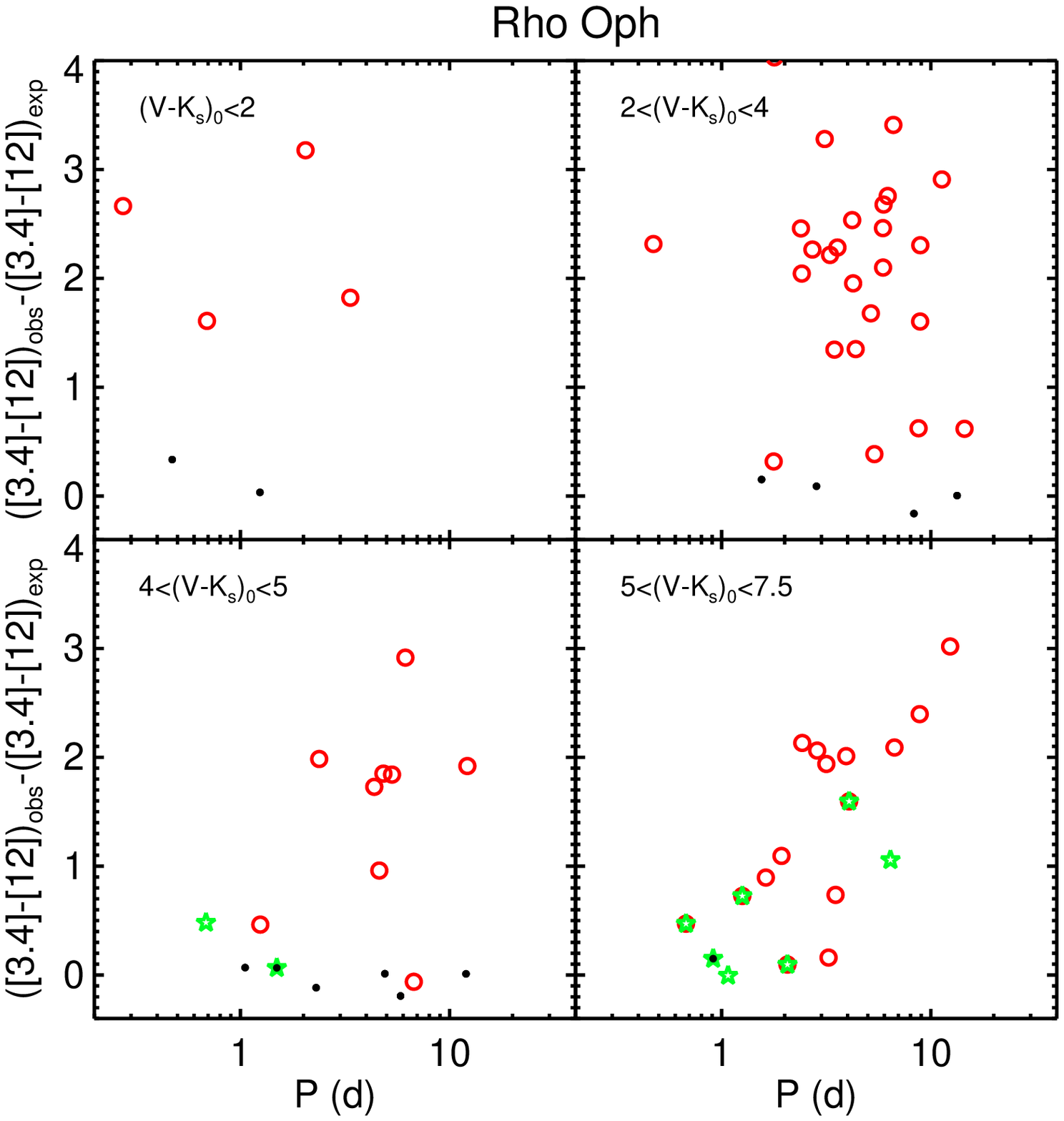}{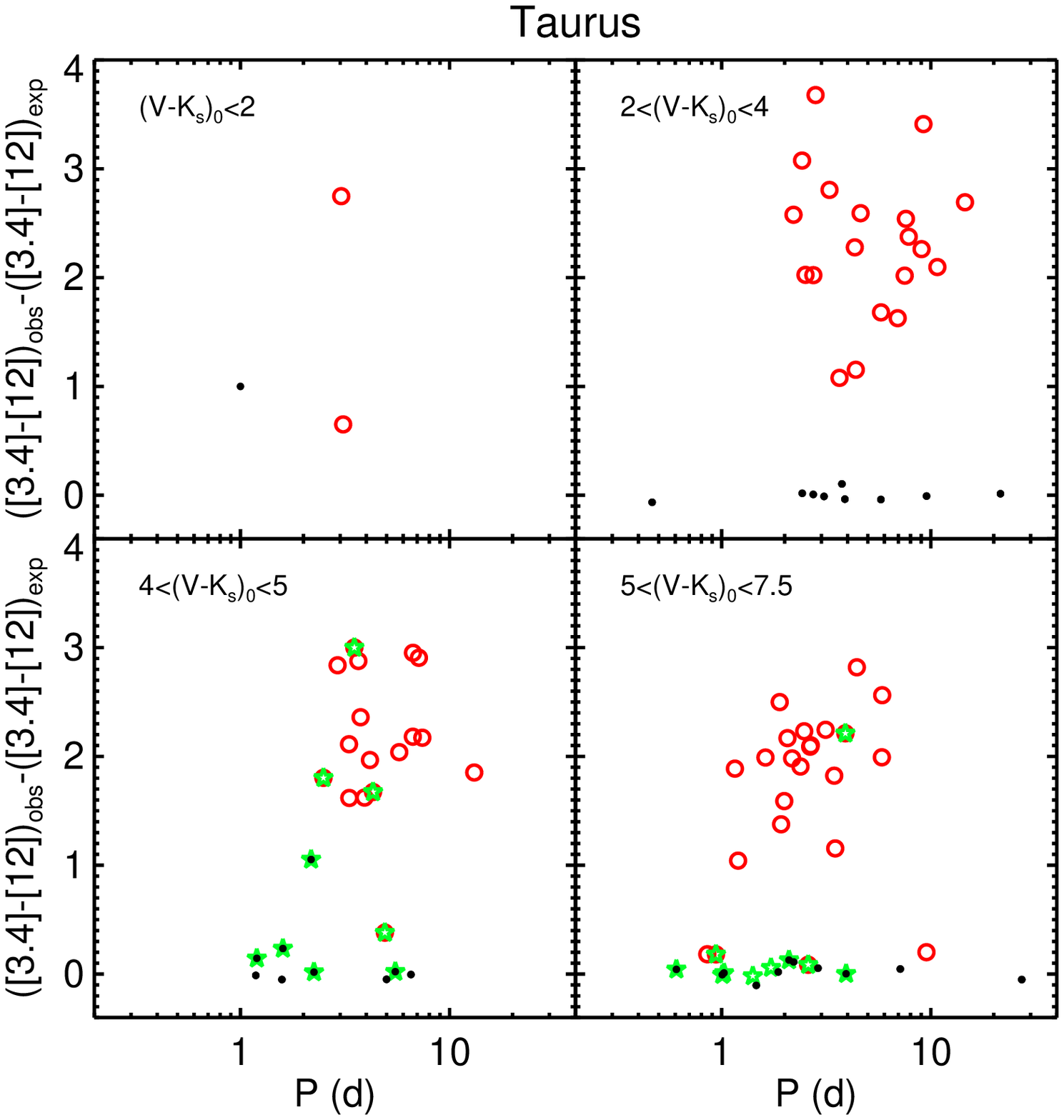}
\caption{Observed [3.4]$-$[12] minus the expected photospheric
[3.4]$-$[12] (see \S\ref{sec:finddisks}) vs.\ $P$ for stars in $\rho$
Oph (left) and Taurus (right; highest confidence members only),
differentiating between the high-confidence disked stars (red, open
circles) and the high-confidence non-disked stars (small black dots). 
M stars are the bottom row; left, 4$<$\vmkz$<$ 5, is roughly M0-M3 and
right, 5$<$\vmkz$<$7.5,  is roughly M4-M5. An additional green star in
these panels denotes that it has more than one period (and is not
tagged a pulsator), e.g., a likely binary; see text.
The relationship in $\rho$ Oph is far
less clear than USco, because there is far more uncertainty in \vmkz\
and there are far fewer stars in comparison to USco. Although
uncertainty in \vmkz\ and far fewer stars also affects Taurus, the
relationship in Taurus more clearly resembles USco than $\rho$ Oph;
Taurus is older than $\rho$ Oph and younger than USco.}
\label{fig:irxrhotau}
\end{figure}

Paper V presented LCs from USco that provide direct evidence for disk
locking, with a striking clumping of disked stars near $\sim$2d,
particularly for the later spectral types. Figure~\ref{fig:irxrhotau}
shows period ($P$) vs.\ IR excess ([3.5]$-$[12]), for $\rho$ Oph
($\sim$1 Myr; data from Paper V) and Taurus ($\sim$3 Myr; just the
highest quality member sample). There are comparable numbers of stars
in these two clusters, and likely comparable uncertainties leading to
\vmkz, though we believe that we have done a slightly better job of
dereddening in Taurus than $\rho$ Oph (see \S\ref{sec:dereddening}).
Uncertainties in $P$ should be low, and reddening should not
significantly affect the [3.5]$-$[12] excess, but reddening will
affect which bins (which panel in the plot) encompass which targets. 

The relationship between $P$ and color seen in USco (Paper V) is not
nearly so convincing in $\rho$ Oph. However, in Taurus, the
relationship is clearer; the disked stars are clumped near $\sim$2d,
and there are very few stars that both rotate more quickly than
$\sim$2d and have disks. Taurus  resembles USco more than $\rho$ Oph,
which makes sense as Taurus is older than $\rho$ Oph (and younger than
USco). Moreover, $\rho$ Oph may be too young for disk locking to
dominate (e.g., Hartmann 2002).  As noted above,
Figure~\ref{fig:vsini} also suggests that the stars with disks rotate
more slowly.

More detailed statistical tests reinforce what is seen by eye. For
USco, all statistical tests find significant differences
(Komolgorov-Smirnov [K-S] and Anderson-Darling for just the period
distributions; 2-D 2-sided K-S for the full distribution of $P$
vs.~\vmkz).  For the other two clusters, the  disked and non-disked
star samples are small enough that the results are just not as clear.
In $\rho$ Oph, the 2-D distributions that are most significantly
different are for 2$<$\vmkz$<$4 (the upper right panel), but there are
only four non-disked stars, so this is not a very robust result. In
Taurus, the $P$ distribution alone for the non-disked sample is close
to that for the disked stars; there is a clumping of the black points
near $\sim$ 3-5d, which is most apparent in the 2$<$\vmkz$<$4 bin (the
upper right panel of Figure~\ref{fig:irxrhotau}). The 1-D $P$
distributions are the most different for the \vmkz$<$2 sample, but
there are very few stars.  The 2-D distributions are significantly
distinct between the disked and non-disked populations in each panel,
but more ambiguously than for USco. (USco 2-D K-S for the
5$<$\vmkz$<$7.5 bin has a probability of $\sim 2 \times 10^{-20}$ that
they come from the same parent population; the same test for the same
Taurus bin is $\sim 0.001$. The same test for the same color range but
for \rop\ is just 0.23.)

We conclude that, within the constraints of the relatively small
number of stars and the uncertainty in  \vmkz, there is some evidence
that disks affect the rotation rate distribution in Taurus. The
distribution of rotation rates vs.\ color in Taurus more closely
resembles that from USco than that from \rop. We note also that there
seems to be few binaries with disks in the lower mass bin in Taurus,
but with the relatively few stars available, it is hard to assess the
significance of this in comparison to the other clusters. 

We also note that all but one of the highest confidence Taurus members
that lack a period from K2 also have disks, supporting the idea that
stochastic contributions from the disk can make periods harder to
find. 

\clearpage

\section{Period-Color Distributions}
\label{sec:rotationdistrib}

In this section, we investigate $P$ as a function of \vmkz, and put
Taurus (and the foreground population) in context with the other
clusters we have studied with K2 data. As in our other K2 rotation
papers,  for stars with more than one period, we have taken the first
period and the measured \vmkz\ as representative of the same star
(likely the primary if it is a multiple); both the assumed \vmkz\ and
first period are listed in Table~\ref{tab:bigperiods}. Even if the
star is a designated multiple identified from additional periods and
position in the CMD, we do not include subsidiary companions
separately in this analysis. Additionally, as before, we assume that
the stars in these three clusters represent snapshots in time of the
same population (c.f., Coker \etal\ 2016).

\subsection{Distribution of $P$ vs.\ $(V-K_s)_0$}
\label{sec:pvmkdistrib}

\begin{figure}[ht]
\epsscale{1.0}
\plotone{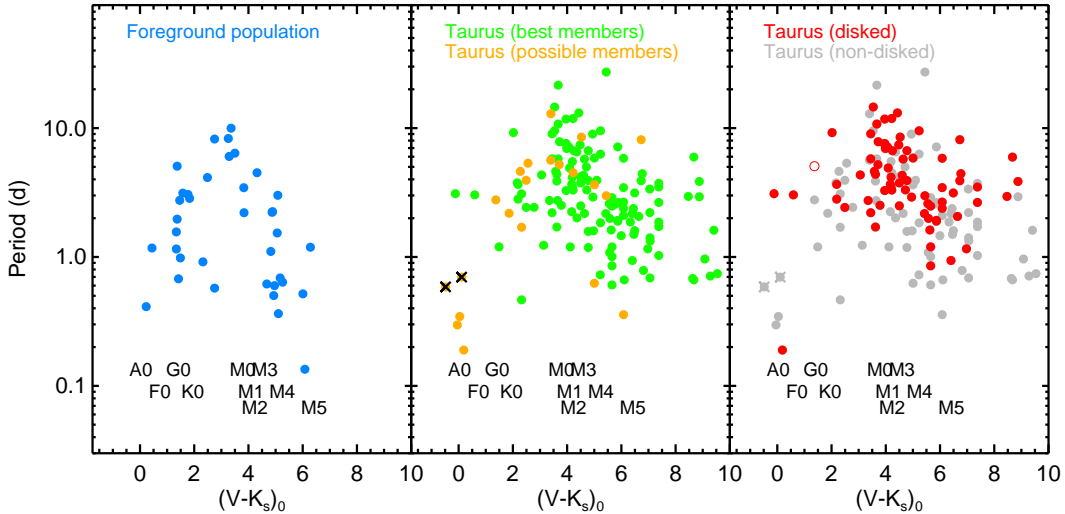}
\caption{$P$ vs.\  \vmkz. Left panel: (blue) foreground population.
Middle panel: green=highest quality Taurus members; orange=lower
confidence Taurus members; an  additional $\times$ denotes stars we
believe to be pulsators.  Right panel: red is disked, grey is
non-disked, for best and possible Taurus members (the single
high-confidence disk among the foreground population is a hollow red
circle).  Approximate spectral types are indicated for reference.  The
slowest rotators are found in the early Ms; fast rotators are found in
the As-Fs-Gs, and in the mid-Ms, the latest type we have here. Stars
from the foreground population on average rotate more quickly than the
younger Taurus members.  Stars with disks on average rotate more
slowly than stars without disks.}
\label{fig:pvmk3a}
\end{figure}

\begin{figure}[ht]
\epsscale{1.0}
\plotone{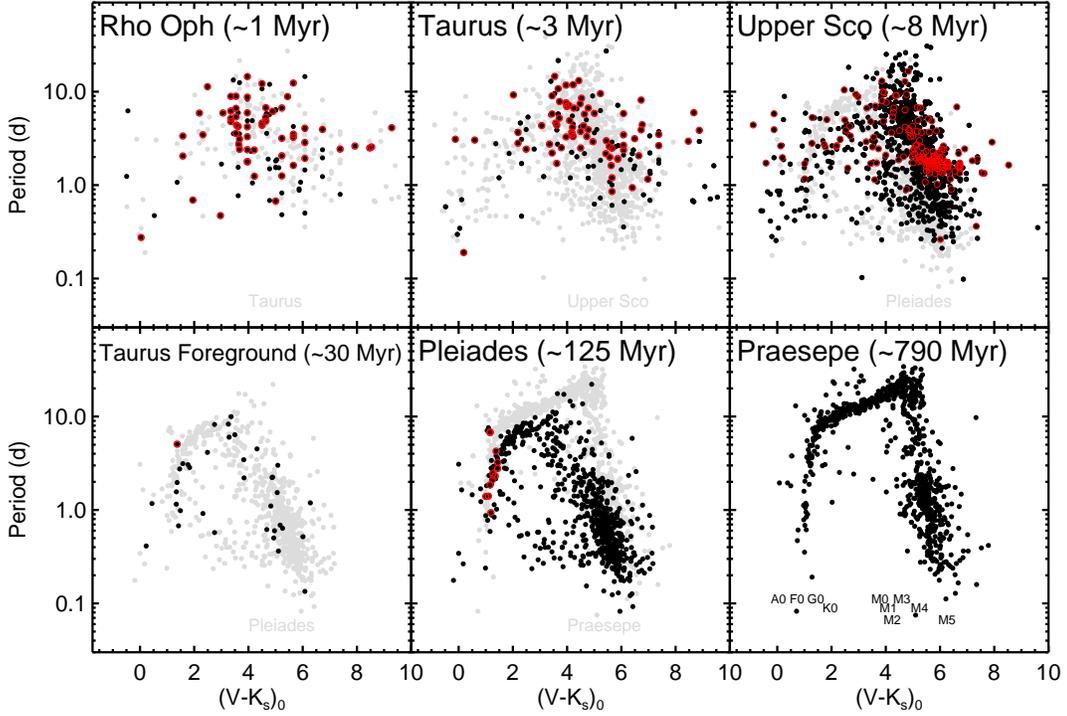}
\caption{$P$ vs.\ \vmkz\ for stars in Taurus and the foreground
population (``Taurus Foreground'') in context with the other clusters
with K2 periods. In each panel, black dots are members, and grey dots
are members from the next panel -- e.g., the Taurus
sample has the USco sample in grey beneath it to aid in comparison
between clusters. (The Pleiades appears for comparison in both USco
and the Taurus foreground because the foreground population is just
too sparse to provide meaningful comparisons to USco.) Additional red
circles denote circumstellar disks; note that Pleiades disks are all
debris disks, while most of the disks in USco and all of those in
Taurus and \rop\ are primordial disks. Upper left: \rop\ ($\sim$1 Myr,
from paper V); upper center: Taurus ($\sim$3 Myr); upper right: USco
($\sim$8 Myr, from paper V); lower left: Taurus foreground ($\sim$30
Myr); lower center: Pleiades (125 Myr, from papers I-III); lower
right: Praesepe ($\sim$790 Myr, paper IV). Approximate spectral types
are indicated for reference in the Praesepe panel. Errors on colors
are conservatively estimated to be $\sim$1 mag for \rop\ and Taurus,
$\sim$0.4 mag for USco and the Taurus foreground, and smaller than the
points for the Pleiades and Praesepe.  Recall that the apparent
quantization of some stars' \vmkz\ (most apparent in \rop\ and some
Taurus) is a result of our dereddening to the expected \vmk\ color for
that spectral type when that method of dereddening was the best option
(see \S\ref{sec:dereddening}). The Taurus distribution has a lot of
scatter, but is starting to resemble the USco distribution. The
(sparse) Taurus foreground distribution is consistent with the
Pleiades distribution. }
\label{fig:multiclusterdiskpaper}
\end{figure}

Figure~\ref{fig:pvmk3a} shows $P$ vs.\  \vmkz. The left panel has the
foreground older population; the others have both the highest
confidence members and the possible members. Overall, the slowest
rotators are found in the early Ms and the fastest rotators are found
in both the highest masses, As and Fs, and in the lowest masses we
have here, mid-Ms.  Despite the many fewer stars available from the
foreground population, it can be seen, on average, that they are
rotating more quickly than the (younger) Taurus members, consistent
with their older ages.  The disked stars are, on average, rotating
more slowly than the disk-free Taurus members, consistent with disk
locking (see Sec.~\ref{sec:disks} above).  The latest M stars in
Taurus are the youngest M stars yet studied with high-quality LCs.

Figure~\ref{fig:multiclusterdiskpaper} puts the $P$ vs.\  \vmkz\ for
Taurus and the foreground population in context with the other young
clusters that have periods we derived from K2 LCs: \rop\ ($\sim$1 Myr,
Paper V), USco ($\sim$8 Myr, Paper V), Pleiades ($\sim$125 Myr, Papers
I-III) \& Praesepe ($\sim$790 Myr, Paper IV).

There are relatively few stars in Taurus (and even fewer foreground
stars) in comparison with the older clusters.  The shortest periods in
all these clusters are on the order of hours and is limited by breakup
(see discussion in Paper V).  The longest periods found in the
youngest clusters are early Ms and are $\sim$10-20 d; there are many
periods longer than $\sim$10d in Praesepe, encompassing K and early M
stars.  At older ages, there are two well defined sequences: rotation
period increasing (slower rotation) to lower mass for FGK stars, and a
second sequence where rotation period decreases sharply (faster
rotation)  as mass decreases among the M dwarfs.   These two trends
(particularly the latter) appear to already be in place by Taurus age 
($\sim$3 Myr).  For higher mass stars (\vmkz$\lesssim$3), the trend is
for higher masses to have faster rotation rates (smaller $P$), and
lower masses to have slower rotation rates, although there are very
few stars available. For low mass (\vmkz$\gtrsim$3)  stars, there are
shorter periods at lower masses.   Such trends are harder to see in
the \rop\ sample, perhaps because of larger uncertainties in the
colors (due to larger and more variable reddening). It does seem to be
there in the Taurus foreground sample as well ($\sim$30 Myr).  There
is an apparent decrease in the range of periods at a given color/mass
as a function of age; this is at least partially (perhaps completely)
due to the larger uncertainties in the inferred colors at younger
ages.   The \vmkz\ colors are more uncertain in the youngest clusters
due to the larger extinction corrections, possibly the influence on
\ks\ from the disk (this is relatively rare in this sample), and
possibly due to variable extinction or accretion (and the
non-simultaneity of the observed or inferred $V$ and measured \ks). 
The Taurus distribution is not as organized as USco, but also not as
disorganized as \rop; the Taurus distribution is starting to resemble
the USco distribution, but still has a lot of scatter, particularly in
the M stars, more like \rop. The Taurus foreground distribution is
sparse indeed, but consistent with the Pleiades distribution. 

The influence of disks can be seen in
Fig.~\ref{fig:multiclusterdiskpaper}; the top row is all primordial
disks, and the 'pile-up' at $\sim$2 days is obvious in USco, but less
so in the younger clusters (see Sec.~\ref{sec:disks} above). The disks
in the Pleiades are all debris disks; the debris disk sample is likely
incomplete among the M stars in Pleiades and Praesepe, since the more
subtle IR excesses indicative of debris disks are harder to find among
the fainter stars.

\section{Linkages to Analysis in Papers I-V}
\label{sec:linkage}

Although the high quality K2 data enable us to derive reliable periods
for active young stars, the presence of disks makes a difference in
the rotation rates and the LC shapes in clusters younger than $\sim$10
Myr. There are some similarities between the LCs we find in Taurus
(and the foreground population) and LCs in the other clusters. In this
section, we provide an assessment of the properties of the Taurus LCs
in the same fashion as our other papers.

\subsection{Amplitudes}

We included in Papers I (Pleiades) and IV (Praesepe) information about
the variability amplitude, where we define amplitude as the difference
in magnitude between the 10th and the 90th percentile of the flux
distribution for a given LC. We did not include such a discussion in
Paper V (USco and \rop) because the systematics in that campaign (C2)
were large and hard to manage (e.g., the fraction of the variability
due to the instrument or due to the star is hard to determine). In
most cases, however, the amplitudes we measured for the `best' LC for
the periodic sources in USco and \rop\ are probably usable in the
ensemble, especially given the relatively large uncertainty in \vmkz.
The instrumental systematics in the Taurus campaigns are much lower
than that in C2, so here again, the amplitudes are probably due
primarily to the astrophysical source. 

\begin{figure}[ht]
\epsscale{1.0}
\includegraphics[width=0.8\linewidth]{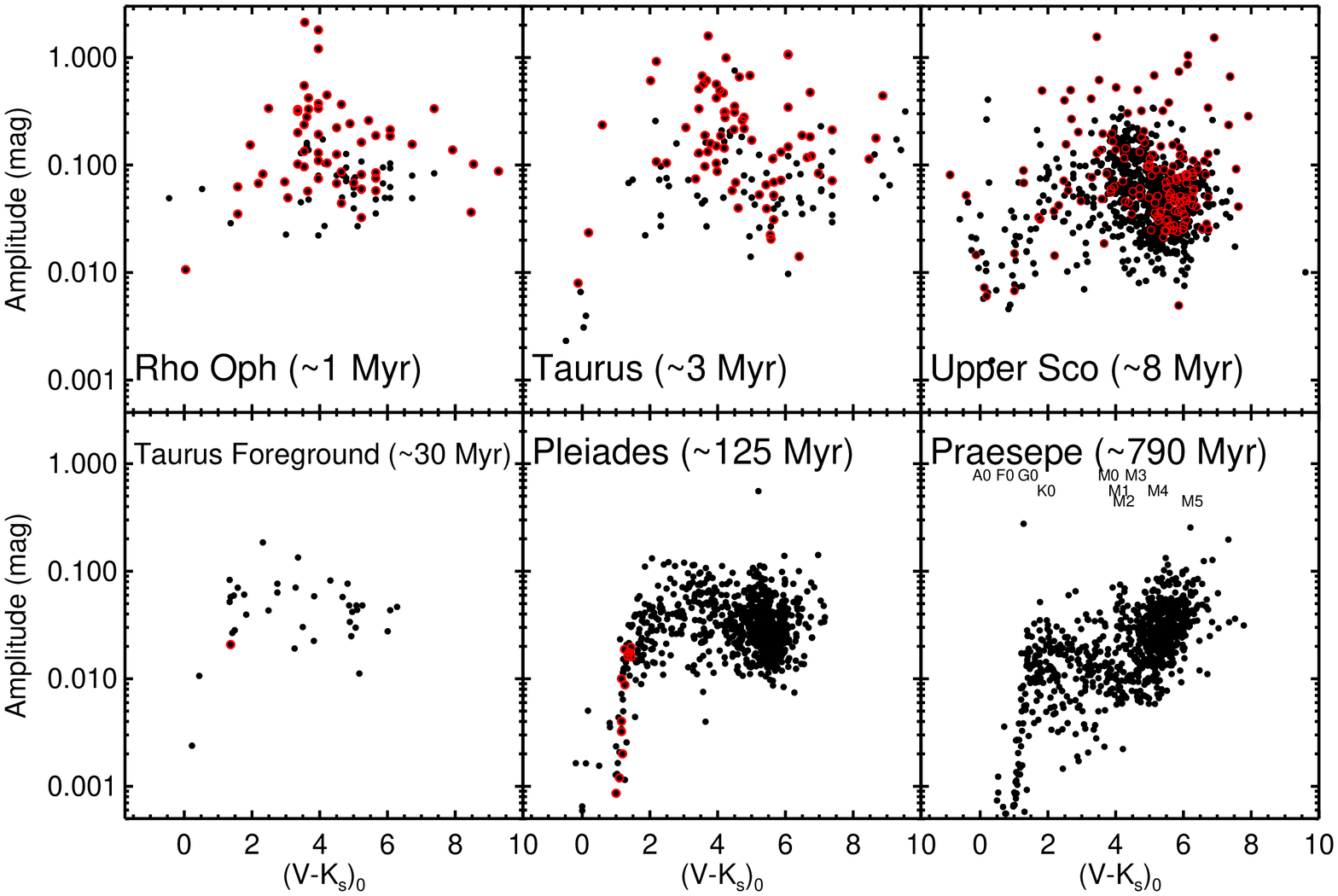}
\\[\baselineskip]
\includegraphics[width=0.8\linewidth]{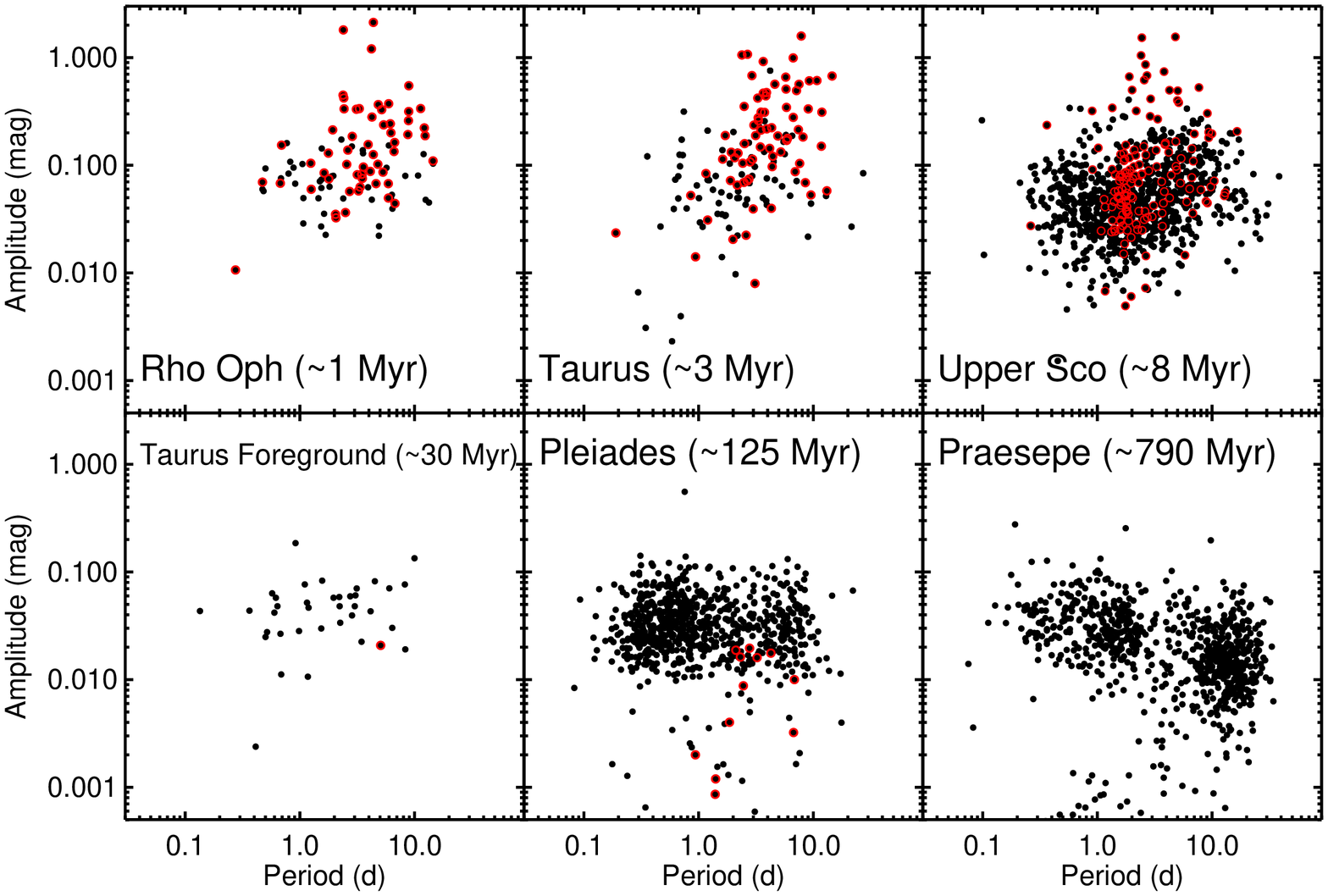}
\caption{Amplitude (10th to 90th percentile) in magnitudes as a
function of \vmkz\ (top) or period in days (bottom) for the same 
clusters as Fig.~\ref{fig:multiclusterdiskpaper}; an additional red
circle indicates a disk. Note that the same $y$-range is used
for each panel; the overall average amplitude decreases with cluster
age, as expected. The average amplitude for stars with disks is larger
than those for stars without disks. There is substructure in the older
clusters that is not obvious in the younger clusters. See the text for
discussion. }
\label{fig:multiclusterampl}
\end{figure}

Figure~\ref{fig:multiclusterampl}  shows the amplitudes of periodic
variability (10th to 90th percentile) as a function of color and
period for our K2 young clusters. As in
Fig.~\ref{fig:multiclusterdiskpaper}, echoes of structures seen in the
older and more populated cluster plots can be seen in the younger
clusters. The average LC amplitude decreases with cluster age.  

Stars with primordial disks have larger amplitudes on average, though 
disked and non-disked stars both have a wide range of amplitudes in
the youngest three clusters (\rop, Taurus, USco).  The variability
amplitudes for the Taurus disked stars primarily measure the larger
amplitude variability due to accretion bursts (``bursters'') or
variable extinction (``dippers'').  The variability amplitudes for the
non-disked stars (and for the low mass stars in the older clusters in
Figure 12) primarily measure non-axisymmetrically distributed spots
as they rotate into and out of our line of sight. 

The earliest spectral type stars sampled here (As and Fs) have lower
amplitudes on average than their same-age counterparts among the Gs,
Ks, and Ms. This makes sense since the A and F stars are expected to
have substantially less fractional spot coverage than the Gs/Ks/Ms at
any given age. 

We expected that longer period rotators might have lower amplitudes on
average, due to lower levels of activity. That is very roughly the
case at Praesepe and perhaps Pleiades ages; it might be the case at
USco age, and seems to not be the case for disk-free younger stars. 

There is obvious sub-structure seen in the amplitude distribution
against color or period in Praesepe (and more subtle echoes of it in
the Pleiades) as a function of \vmkz, some of which may be selection
effects in that the longest periods may require larger amplitudes for
detection (Paper IV).  The sub-structure seen in the amplitude
distribution in the Pleiades and Praesepe is not obvious in the
clusters $<$100 Myr -- perhaps the behavior is not yet organized
enough, or there are not enough stars for comparison.  That lack of
structure could be explained in the \vmkz\ plot by invoking the likely
large uncertainties in \vmkz\ for the young cluters, but the
uncertainty in $P$ is likely small, so the relative lack of
substructure in the distributions as a function of $P$ for the younger
stars is likely real, though greatly complicated by the range of
masses included at any given $P$ (or amplitude).

\clearpage

\subsection{LC and Periodogram Categories}
\label{sec:LCPcats}

In papers II, IV, and V, we classified the LC and periodogram
morphologies for the Pleiades, Praesepe, USco, and \rop; see these
references for discussion of the classes (e.g., double dip, scallop,
etc.). Table~\ref{tab:multipstats} includes statistics on these
classifications for all the clusters, now including Taurus. 

A high fraction of the Taurus stars are periodic, 83\%. As for nearly
all of the other clusters, $\sim$70\% of the periodic Taurus members
have only one period; about 30\% of the periodic sample has at least
two real periods. The Taurus foreground has different proportions, but
is likely biased/incomplete 

There are some aspects of the periodic light curve morphologies which
are common between the clusters, and others which show signs of
evolution with stellar age.  Specifically:

\begin{itemize} 

\item The fraction of double-dip stars falls substantially between
Praesepe and \rop. Assuming that the rate as which we find double-dip
characteristics increases as rotation period decreases (e.g., Basri
\& Nguyen 2018), this makes sense since Praesepe is, on average,
rotating much more slowly than \rop. 

\item LCs of the scallop shell morphology (and related categories)
continue to be rare but much more common among the younger clusters.
None have disks.  We note that Zhan \etal\ (2019) find scallop shell
stars at $\sim$40 Myr, but there are very few by $\sim$125 Myr,
Pleiades age.

\item LC categories suggestive of complicated spot occurence and evolution
(beater, complex peak) seem harder to find in clusters with a higher
disk fraction. In contrast, shape changers (which could arise from
spot/spot group emergence/evolution or from disk influence) are more
common in clusters with a higher disk fraction.

\item Resolved distant peaks (which are most likely binaries, either real or
apparent) occur at comparable rates across all clusters. Pulsators are
rare in all clusters, which is unsurprising, at least in the stellar
mass range where our studies are focused.

\item Dippers and bursters only occur in clusters that still have primordial
disks, consistent with our physical interpretation of these phenomena
as interactions with gas/dust disks. The vast majority of dippers and
bursters are also disked stars; a few have little or no IR excess,
though they could still have gas disks depleted in dust. 

\end{itemize}

\floattable
\begin{deluxetable}{lcccccccccccccccccccccccc}
\tabletypesize{\scriptsize}
\floattable
\rotate
\tablecaption{Star/LC/Periodogram
Categories\tablenotemark{a} \label{tab:countclasses}\label{tab:multipstats}.  }
\tablewidth{0pt}
\tablehead{\colhead{Category} 
& \multicolumn{3}{c}{{\bf Praesepe}} 
& \multicolumn{4}{c}{{\bf Pleiades}} 
& \multicolumn{3}{c}{{\bf Tau F'gnd }}
& \multicolumn{4}{c}{{\bf USco}} 
& \multicolumn{4}{c}{{\bf Taurus }}
& \multicolumn{4}{c}{{\bf \rop\ }} \\ 
& \colhead{(1)\tablenotemark{b}} & \colhead{(2)\tablenotemark{c}} & \colhead{(3)\tablenotemark{d}}
& \colhead{(1)\tablenotemark{b}} & \colhead{(2)\tablenotemark{c}} & \colhead{(3)\tablenotemark{d}}& \colhead{(4)\tablenotemark{e}} 
& \colhead{(1)\tablenotemark{b}} & \colhead{(2)\tablenotemark{c}} & \colhead{(3)\tablenotemark{d}} 
& \colhead{(1)\tablenotemark{b}} & \colhead{(2)\tablenotemark{c}} & \colhead{(3)\tablenotemark{d}}& \colhead{(4)\tablenotemark{e}} 
& \colhead{(1)\tablenotemark{b}} & \colhead{(2)\tablenotemark{c}} & \colhead{(3)\tablenotemark{d}}& \colhead{(4)\tablenotemark{e}} 
& \colhead{(1)\tablenotemark{b}} & \colhead{(2)\tablenotemark{c}} & \colhead{(3)\tablenotemark{d}}& \colhead{(4)\tablenotemark{e}}}
\startdata
Member LCs              &  938 & 1.00 & \ldots &   826 & 1.00 & \ldots & \ldots &   38 & 1.00 & \ldots &  1136 & 1.00 & \ldots & \ldots &  179 & 1.00 & \ldots & \ldots &  174 & 1.00 & \ldots &\ldots \\
Periodic                & 819 & 0.87 & 1.00 &  759 & 0.92 & 1.00 & 1.00&  37 & 0.97 & 1.00 &  974 & 0.86 & 1.00 & 0.81& 148 & 0.83 & 1.00 & 0.71& 106 & 0.61 & 1.00 & 0.73\\
Single period           & 674 & 0.72 & 0.82 &  598 & 0.72 & 0.79 & 0.50&  25 & 0.66 & 0.68 &  756 & 0.67 & 0.78 & 0.68& 108 & 0.60 & 0.73 & 0.58&  85 & 0.49 & 0.80 & 0.61\\
Multi-period            & 145 & 0.15 & 0.18 &  161 & 0.19 & 0.21 & 0.50&  12 & 0.32 & 0.32 &  218 & 0.19 & 0.22 & 0.12&  40 & 0.22 & 0.27 & 0.13&  21 & 0.12 & 0.20 & 0.11\\
\hline
Double-dip              & 163 & 0.17 & 0.20 &  107 & 0.13 & 0.14 & 0.25&   6 & 0.16 & 0.16 &  133 & 0.12 & 0.14 & 0.09&  12 & 0.07 & 0.08 & 0.05&   6 & 0.03 & 0.06 & 0.04\\
Moving double-dip       & 121 & 0.13 & 0.15 &   31 & 0.04 & 0.04 & 0.17&   5 & 0.13 & 0.14 &   32 & 0.03 & 0.03 & 0.02&   5 & 0.03 & 0.03 & 0.01&\ldots&\ldots&\ldots&\ldots\\
Shape changer           & 297 & 0.32 & 0.36 &  114 & 0.14 & 0.15 & 0.25&  16 & 0.42 & 0.43 &  277 & 0.24 & 0.28 & 0.43&  83 & 0.46 & 0.55 & 0.58&  47 & 0.27 & 0.44 & 0.45\\
Scallop/clouds?\tablenotemark{f} &\ldots&\ldots&\ldots&   5 & 0.01 & 0.01 & \ldots &   1 & 0.03 & 0.03 &   28 & 0.02 & 0.03 & \ldots&   6 & 0.03 & 0.04 & 0.00&\ldots&\ldots&\ldots&\ldots\\
Beater                  &  77 & 0.08 & 0.09 &  136 & 0.16 & 0.18 & 0.75&   7 & 0.18 & 0.19 &  107 & 0.09 & 0.11 & 0.03&  18 & 0.10 & 0.12 & 0.05&  10 & 0.06 & 0.09 & 0.06\\
Complex peak            &  68 & 0.07 & 0.08 &   89 & 0.11 & 0.12 & 0.42&   2 & 0.05 & 0.05 &    8 & 0.01 & 0.01 & \ldots&   6 & 0.03 & 0.04 & 0.02&\ldots&\ldots&\ldots&\ldots\\
Resolved, close peaks   &  71 & 0.08 & 0.09 &  127 & 0.15 & 0.17 & 0.42&   9 & 0.24 & 0.24 &  151 & 0.13 & 0.16 & 0.08&  25 & 0.14 & 0.17 & 0.07&  12 & 0.07 & 0.11 & 0.04\\
Resolved, distant peaks &  77 & 0.08 & 0.09 &   39 & 0.05 & 0.05 & 0.17&   5 & 0.13 & 0.14 &   85 & 0.07 & 0.09 & 0.05&  18 & 0.10 & 0.12 & 0.06&   9 & 0.05 & 0.08 & 0.08\\
Pulsator                &  17 & 0.02 & 0.02 &    8 & 0.01 & 0.01 & 0.00&\ldots&\ldots&\ldots&  13 & 0.01 & 0.01 & 0.01&   2 & 0.01 & 0.01 & \ldots&\ldots&\ldots&\ldots&\ldots\\
\hline
Disk                    &\ldots&\ldots&\ldots&12&0.01 & 0.02 & 1.00 &\ldots&\ldots&\ldots&208&0.18 & 0.17 & 1.00 &101&0.56 & 0.49 & 1.00 &80&0.46 & 0.55 & 1.00 \\
Dipper\tablenotemark{g} &\ldots&\ldots&\ldots&\ldots&\ldots&\ldots&\ldots&\ldots&\ldots&\ldots&  66 & 0.06 & 0.06 & 0.31&  21 & 0.12 & 0.12 & 0.21&  17 & 0.10 & 0.15 & 0.20\\
Burster\tablenotemark{g}&\ldots&\ldots&\ldots&\ldots&\ldots&\ldots&\ldots&\ldots&\ldots&\ldots&  24 & 0.02 & 0.01 & 0.11&  25 & 0.14 & 0.11 & 0.25&  13 & 0.07 & 0.08 & 0.16\\
\enddata
\tablenotetext{a}{This table has been updated to reflect slightly new
classifications in a few cases for stars in clusters we have already
published, and one error -- the total sample in the Pleiades had been
listed in Paper I/II as only the highest quality members, but all the
other numbers lumped the highest quality and `ok' members together.}
\tablenotetext{b}{Total number of stars in the (sub)sample; for
Taurus, it is the combination of the highest quality and possible
members. Examples for Praesepe: there are 938 members, and 819
periodic members. }
\tablenotetext{c}{Sample fraction (fraction of the given type), out of
the members. Examples for Praesepe: 87\% of the members are periodic;
72\% of the members are single-period. }
\tablenotetext{d}{Periodic sample fraction (fraction of the given
type), out of the periodic members. Examples for Praesepe: 82\% of the
periodic members are single-period; 18\% of the periodic members are
multi-periodic.}
\tablenotetext{e}{Disked sample fraction (fraction of the given type),
out of the disked members. Note that this column does not appear for
those clusters without disks. Examples for Pleiades: 100\% of the
disked member stars are periodic; 50\% of the
disked member stars are single-period; 25\% of the disked member stars
are double-dip.}
\tablenotetext{f}{Described in papers I-V sometimes as ``orbiting
clouds?'', Stauffer \etal\ (2017, 2018a) categorized scallops,
persistent flux dips, and transient flux dip stars. Here we simply use
the term `scallop' to include all three light curve categories (Stauffer
\etal\ 2017, 2018a).}
\tablenotetext{g}{Some objects are tagged burster or dipper, but not
periodic. The fraction of the periodic sample that is burster or
dipper is correct, for the periodic subsample.}
\end{deluxetable}


\begin{figure}[ht]
\epsscale{0.8}
\plotone{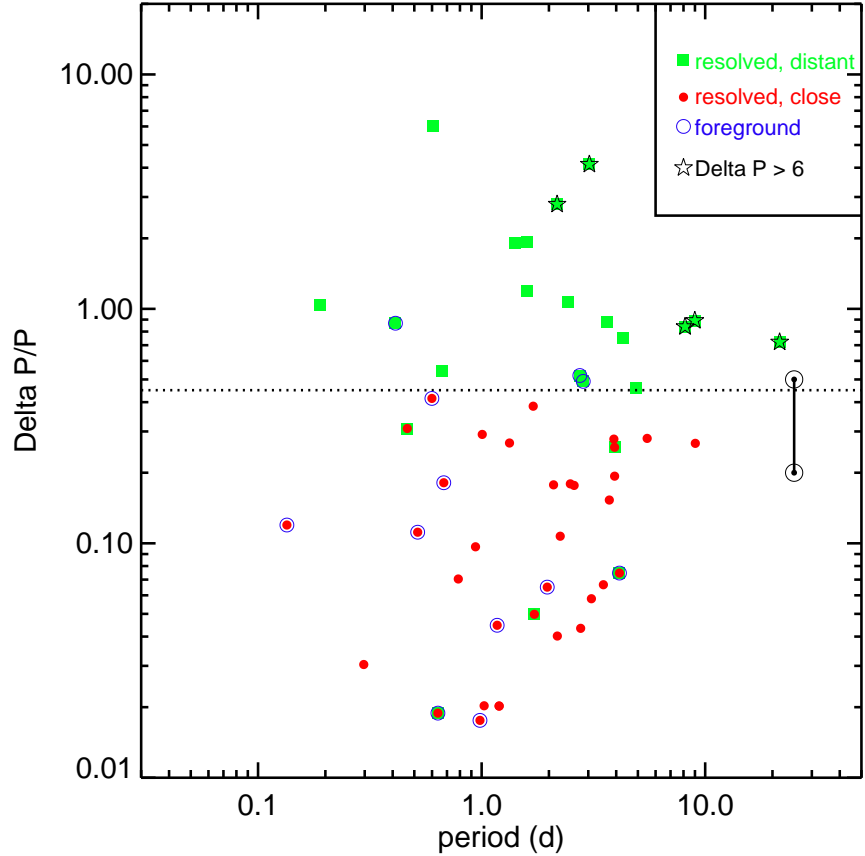}
\caption{Plot of $\Delta P/P_1$ vs.~$P$ for stars that are
multi-periodic and with resolved distant peaks
(green squares), and resolved close peaks (red dots) for
high-confidence and possible members. An additional black star
indicates that $|\Delta P|>$6 d. An additional blue circle means the
star is from the older foreground population   The range of
possible values for the  Sun is included for reference ($\odot$);  if
one takes as $\Delta P$ the range of periods measured where sunspots
occur, $\Delta P/P_1 \sim 0.1-0.2$, but if one takes the full range of
$\Delta P$, equator to pole, $\Delta P/P_1 \sim 0.5$. The dotted line
is at $\Delta P/P_1$=0.45 and denotes the boundary between close and
distant resolved peaks.  There are far fewer structures in this
diagram as compared to that for the Pleiades; it's possible that there
are not enough complete cycles in the $\sim$70d campaign for
extraction of clear multiple periods, or the disk influence on the LCs
complicates derivation of the periods.}
\label{fig:deltap}
\end{figure}

\clearpage

\subsection{$\Delta P$ distributions}
\label{sec:deltaP}

As for the other clusters, we calculated the $\Delta P/P_{\rm rot}$
metric for stars with resolved multi-period peaks; see
Figure~\ref{fig:deltap}. This metric $\Delta P/P_{\rm rot}$ is defined
by taking the closest peak to the $P_{\rm rot}$, subtracting the
smaller from the larger, and dividing by the $P_{\rm rot}$. For
well-separated peaks and/or M stars (see Paper V and Stauffer \etal\
2018b), it is very likely that the two periods are two independent
periods of two components of a binary system. 

The distribution in Taurus, like that for USco/\rop\ (paper V) and
Praesepe (paper IV), has relatively little structure as compared to
the Pleiades (paper II). Here, as noted in paper V, the $\sim$70 d K2
campaigns are not quite long enough to capture multiple complete cycles
in order to be able to resolve two close periods. If the phasing of the two
periods is favorable, we can resolve the two periods, but if the
phasing is not `lucky,' $\sim$70 d is not long enough to distinguish
the periods.

\section{Summary and Conclusions}
\label{sec:concl}

We have presented the K2 LCs and periods derived from them for the
Taurus star forming region. We have LCs for about 30\% of the highest
confidence Taurus members in the literature; most of the Taurus
members are North of the region observed by K2. However, because the
K2 region extends over a larger area, it enables study not only of
bona fide Taurus molecular cloud members, but also of a somewhat
older, more dispersed population of candidate young stars. 

We followed the methodology used in our earlier papers I-V and find,
again, that cluster membership is one of the most difficult parts of
our analysis. We distributed all plausible candidate members of Taurus
that were observed by K2  amongst four bins -- highest quality
members, possible members, a population of stars in front of the
Taurus members (``foreground'') that are likely about 3 times older
than Taurus members, and NM.  We retain for further analysis a sample
consisting of the highest quality members and the possible members.
Dereddening is also challenging and likely contributes 
uncertainty to the \vmkz\ colors. We used IR data to identify stars
with disks. We classified the LCs according to the classes we
identified in papers I-V. We retained up to four periods for each star.

We find slightly more than 80\% of the Taurus members to be periodic;
the foreground population is highly biased/incomplete, but
nearly all of the foreground population is also periodic.

We find that the periods from our analysis are generally well-matched
to the (substantial) previous literature, except for possibly the KELT
analysis (Rodriguez \etal\ 2017b). Our periods are also consistent
with the available literature \vsini\ values, and our analysis
suggests that there is no substantial bias in our sample introduced by
our inability to get periods from stars with a high level of
stochastic contributions (from, e.g., a disk or accretion). 

Despite the relatively paucity of stars with K2 LCs and derived
periods in Taurus (as compared to USco), similar relationships between
rotation rate and disk presence are found in Taurus. Especially for
the M stars, the periods are clumped near $\sim$2 d, and there are
very few stars that have disks and rotate faster than $\sim$2 d.

The latest M stars in Taurus are the youngest M stars yet studied with
high-quality LCs, extending into the brown dwarf regime. 

The overall distribution of rotation rates as a function of color (as
a proxy for mass) in Taurus, and in the foreground population, is similar
to that found in other clusters: the slowest rotators are in the early
Ms, and both the higher masses and lower masses are the fastest
rotators. The member distribution looks similar to that from \rop\ and
USco (paper V), suggesting that the distribution (particularly for M
stars) is already in place by Taurus age ($\sim$3 Myr). The slightly
older ($\sim$30 Myr) foreground population has very few stars, but is
consistent with the USco and Pleiades distributions. 

The LC amplitude distribution as a function of \vmkz\ or period has
trends and substructure, not all of which is easy to interpret. The
overall amplitude goes down as clusters age. Higher mass stars have
lower amplitudes than lower mass stars; however there is a selection
effect in that amplitudes of the faintest stars have to be larger to
be detected.  Stars with primordial disks have, on average, larger
amplitudes, though the mechanisms driving the variability are
different than for stars without disks. Interpretation of additional
structure in these distributions is much less clear. 

As for many of the other young clusters observed by K2, about 70\% of
the periodic members have just one period detected, and about 30\% of
the periodic members have more than one period. Scallop shell/orbiting
dust cloud stars are still rare; there are just 7 in the  sample
discussed here (1 from the foreground, 6 from Taurus).  LC categories
that we have interpreted as spot/spot group changes
(emergence/evolution), namely beaters and complex peaks, seem harder
to find in clusters with many primordial disks. Shape changers, which
could be from spot/spot group changes or from disk interactions and/or
accretion, are more common in clusters with a higher disk fraction.
Dippers and bursters are only found in clusters with primordial disks;
most of these specific objects have obvious dusty disks. 

Cody \etal\ (2020 in prep) will explore the range of light curve (LC)
properties of the disked members of Taurus. Roggero \etal\ (2020 in
prep) will explore the dipper population in Taurus.

\acknowledgments

Some of the data presented in this paper were obtained from the
Mikulski Archive for Space Telescopes (MAST). Support for MAST for
non-HST data is provided by the NASA Office of Space Science via grant
NNX09AF08G and by other grants and contracts. This paper includes data
collected by the Kepler mission. Funding for the Kepler mission is
provided by the NASA Science Mission directorate. 
This research has made use of the NASA/IPAC Infrared Science Archive
(IRSA), which is operated by the Jet Propulsion Laboratory, California
Institute of Technology, under contract with the National Aeronautics
and Space Administration.    This research has made use of NASA's
Astrophysics Data System (ADS) Abstract Service, and of the SIMBAD
database, operated at CDS, Strasbourg, France.  This research has made
use of data products from the Two Micron All-Sky Survey (2MASS), which
is a joint project of the University of Massachusetts and the Infrared
Processing and Analysis Center, funded by the National Aeronautics and
Space Administration and the National Science Foundation. The 2MASS
data are served by the NASA/IPAC Infrared Science Archive, which is
operated by the Jet Propulsion Laboratory, California Institute of
Technology, under contract with the National Aeronautics and Space
Administration. This publication makes use of data products from the
Wide-field Infrared Survey Explorer, which is a joint project of the
University of California, Los Angeles, and the Jet Propulsion
Laboratory/California Institute of Technology, funded by the National
Aeronautics and Space Administration. 

\facility{K2, 2MASS, WISE, IRSA, Exoplanet Archive}

\clearpage

\appendix

\section{Membership}
\label{app:membership}

\subsection{Members}
\label{app:members}

Despite the fact that Taurus is close ($\sim$140 pc) and has
been studied for decades, membership is still controversial.  Recent
literature combines years of prior data collection with the Gaia DR2
data in attempts to create a complete list of members (e.g., Esplin \&
Luhman 2019; Luhman 2018; Galli \etal\ 2019; Luhman \etal\ 2017; and
references therein). Numerous approaches have sought a dispersed
population of young stars (e.g., Kraus \etal\ 2017; Rebull \etal\
2011, 2010; Slesnick \etal\ 2006; Wichmann \etal\ 1996; and references
therein). The net result is persistently confusing literature, with a
subset of stars that most authors agree must be members, stars where
membership evidence is mixed, and still other stars where evidence is
both mixed and controversial, particularly those having several
indications of youth but with distinct motions and/or distances that
are apparently inconsistent with membership. 

In the context of this paper, we needed to find the sample of member
stars that also had K2 LCs. It is important to note three important
selection effects. First, K2 LCs were only downloaded for targets
submitted in advance, so completeness is limited by the submitted list
of targets. Second, the K2 observations did not cover all of the sky
encompassing the least controversial Taurus members (see
Fig.~\ref{fig:where}, in addition to discussion below). Third, since
it observed in optical bands, K2 effectively selected against those
Taurus members that are so embedded as to attenuate the optical data
(see discussion below and Sec.~\ref{sec:finddisks}). 

Based on our experience with the other K2 clusters we have analyzed
(Papers I-V), we knew that we had to start with an overly generous
list of possible members; by starting with an expansive set, we could
analyze all of the LCs in the same way, at the same time (thus
reducing biases in the analysis). Subsequent membership modifications
implemented late in the process then become trivial to implement.  

We assembled a list of 851 EPIC-numbered targets with LCs (e.g.,
targets submitted in advance that had actually been observed) that
could be members of Taurus with data in either of the Taurus-relevant
K2 campaigns (C4 and C13). These targets originated from any of 64
different K2 GO programs that mention in their abstract studying young
stars or members of Taurus. These programs did not necessarily do
complete and thorough vetting of possible targets prior to target
submission (we note both C4 and C13 predate Gaia DR2); observers just
included targets that could be Taurus members, so we likewise included
them in our intitial target list assembly.

We proceeded with the analysis of those $\sim$850 LCs in
parallel with membership assessment efforts. For most of these targets
(see Sec.~\ref{sec:litphotom} and Table~\ref{tab:summarystats}), there
are Gaia DR2 and 2MASS data. We explored a variety of parameter spaces
such as vector point diagrams and color-magnitude diagrams. While this
process was underway, Luhman (2018) was published, with similar
subgroups as we had identified, but with a much more complete analysis
than we had accomplished to that point. Subsequently, Esplin \& Luhman
(2019) and Galli \etal\ (2019) were published, again presenting lists
of members based primarily on Gaia DR2 data.  These recent analyses
include all of the least controversial Taurus members. We made the
decision to adopt the intersection of those targets with K2 LCs and
those in Esplin \& Luhman (2019) as the {\em highest quality
members.}

Note that the regions of sky encompassed by Luhman (2018),
Esplin \& Luhman (2019), and Galli \etal\ (2019) are significantly
different than the region of sky observed by K2 in the two campaigns
including Taurus members (Fig.~\ref{fig:where}). The K2 observations
cover about 30\% of the Taurus members from Esplin \& Luhman (2019),
and the K2 observations also encompass a much larger region of the sky
to the south of where the membership work focused.  Moreover, the
Esplin \& Luhman (2019) member list seems to have very few higher mass
G and K stars\footnote{A similar G star  deficit was noted also by
Tokovinin \& Brice\~no (2020) in the context of the USco membership
determined by Luhman and collaborators.}. Since that mass range proved
interesting for the rotation rate distribution in Papers I-V, we
wanted to be sure that we had that region of parameter space
well-represented here. 

As a result of the larger areal coverage and the G star deficit, we
sought to reassess membership for stars with K2 LCs that were not
already in the highest quality members, in particular stars meeting
those criteria. We investigated the literature for each possible
additional target in question, as well as the Gaia DR2 data. In most
cases, a star had several or many papers in the literature citing it
as a suspected Taurus member and it had DR2 data consistent with
belonging to Taurus, or it had very few references suggesting
membership and usually DR2 data inconsistent with membership. The {\em
less likely (lower-confidence) members} include targets added by our
own additional star-by-star inspections, including a few from Galli
\etal\ (2019) or Kraus \etal\ (2017) that weren't in Esplin \& Luhman
(2019).

In this process, we realized that there was a subset of stars
that had indications of youth but distances and/or motions
incompatible with Taurus membership; these include those
`controversial' members mentioned above. Many of these targets are
listed in Luhman (2018) as Group 29 members (see also Oh \etal\ 2017).
Some appear in the literature as possible members of Mu Tau (Gagne
\etal\ in prep, Liu \etal\ 2020), or possibly Cas-Tau (e.g., Hartmann 
\etal\ 1991, de Zeeuw \etal\ 1999, Luhman 2018; also see appendix of
David \etal\ 2018). Analysis to date of this population in the
literature suggest that they are $\sim$30 Myr (e.g., Kraus \etal\
2017). Our list of {\em Taurus Foreground} stars are primarily those
Group 29 members, plus targets added by our own additional
star-by-star inspections that might very well be Mu Tau or Cas-Tau
members. (See also Stauffer \etal\ in prep.) 

Much of the supporting non-survey data available in the literature
(Sec.~\ref{sec:litphotom}) is biased towards those least controversial
members, and those targets nearby in projected distance. For example,
a high fraction of the highest confidence members have a Spitzer
and/or Herschel counterpart (Sec.~\ref{sec:litphotom} and
\ref{sec:finddisks}; Table~\ref{tab:summarystats}) because those
pointed missions targeted the most well-known members. Circularly, the
most reliable members also are more likely to have an IR excess, and
thus be detected in various IR data sets.  Similarly, we expect
members to be young and significantly spotted and therefore periodic
more often, and that is the case (Sec.~\ref{sec:periods}).

\subsection{Non-Members}
\label{app:nm}

After extracting the members (of any sort) out of the $\sim$850
in our initial sample, we are left with $\sim$630 nominal NM.  Among
this nominally NM sample, there is an anomalously high fraction of
apparent IR excesses and an anomalously high  periodic fraction
(Table~\ref{tab:summarystats}). Looking at the distribution of Gaia
distances from Bailer-Jones \etal\ (2018), many are nominally at
$\sim$140 pc, more than would be expected for a randomly selected set
of field stars. We suspect that we have been somewhat conservative in
identifying members, and that some (perhaps several) legitimate
members are left in the NM sample. Moreover, the set of stars with K2
LCs is biased towards things appearing in the literature as young,
which makes the entire K2 sample different than any random selection
of stars.

For these NM, we did much of the same analysis as for our
members, but they were largely ignored in the discussion in the rest
of the paper. We include results from these additional stars here,
should they be of use to future investigators.  We provide the NM in
Table~\ref{tab:bignm}, which has the same contents as in
Table~\ref{tab:bigperiods} above (save for the column on cluster
membership). Note that the Gaia DR2 ID is provided for easy matching
to parallaxes and proper motions. We focused on analysis of the
members, and as such some of the NM may not have been vetted as
closely as the member sample.

\floattable
\begin{deluxetable}{ccp{13cm}}
\tabletypesize{\scriptsize}
\tablecaption{Contents of Table: Periods and Supporting Data for
Taurus Likely Non-Members with K2 Light Curves\label{tab:bignm}}
\tablewidth{0pt}
\tablehead{\colhead{Number} & \colhead{Label} & \colhead{Contents}}
\startdata
1 & EPIC & Number in the Ecliptic Plane Input Catalog (EPIC) for K2\\
2 & coord & Right ascension and declination (J2000) for target \\
3 & othername & Alternate name for target \\
4 & gaiaid & Gaia DR2 ID \\
5 & Vmag & V magnitude (in Vega mags), if observed\\
6 & Kmag & \ks\ magnitude (in Vega mags), if observed\\
7 & vmk-obs & \vmk, as directly observed (in Vega mags), if $V$ and \ks\ exist\\
8 & vmk-used & \vmk\ used, in Vega mags (observed or inferred; see text)\\
9 & evmk & $E(V-K_s)$ adopted for this star (in mags; see \S~\ref{sec:dereddening}) \\
10 & Kmag0 & dereddened $K_{s,0}$ magnitude (in Vega mags), as inferred (see \S\ref{sec:dereddening})\\
11 & vmk0 & $(V-K_s)_0$, dereddened $V-K_s$ (in Vega mags), as inferred (see \S~\ref{sec:dereddening}; rounded to nearest 0.1 to emphasize the relatively low accuracy)\\
12 & uncertaintycode & two digit code denoting origin of \vmk\ and \vmkz\
(see \S\ref{sec:litphotom} and \ref{sec:dereddening}):
First digit (origin of \vmk): 
1=$V$ measured directly from the literature (including SIMBAD) and $K_s$ from 2MASS; 
2=$V$ from APASS and $K_s$ from 2MASS;
3=\vmk\ inferred from Gaia $g$ and $K_s$ from 2MASS (see \S\ref{sec:litphotom});
4=\vmk\ inferred from Pan-STARRS1 $g$ and $K_s$ from 2MASS (see \S\ref{sec:litphotom});
5=\vmk\ inferred from membership work (see \S\ref{sec:membership}; rare); 
6=$V$ inferred from well-populated optical SED and $K_s$ from 2MASS (see \S\ref{sec:litphotom});
-9= no measure of \vmk.
Second digit (origin of $E(V-K_s)$ leading to \vmkz): 
1=dereddening from $JHK_s$ diagram (see \S\ref{sec:dereddening});
2=dereddening back to \vmkz\ expected for spectral type;
3=used median $E(V-K_s)$=0.7 (see \S\ref{sec:dereddening});
-9= no measure of  $E(V-K_s)$ \\
13 & P1 & Primary period, in days (taken to be rotation period in cases where there is $>$ 1 period)\\
14 & P2 & Secondary period, in days\\
15 & P3 & Tertiary period, in days\\
16 & P4 & Quaternary period, in days\\
17 & IRexcess & Whether an IR excess is present or not (see \S\ref{sec:disks})\\
18 & IRexcessStart & Minimum wavelength at which the IR excess is detected or the limit of our knowledge
of where there is no excess (see \S\ref{sec:disks}) \\
19 & slope & Slope of SED fit to all available detections between 2 and 25 \mum\ \\
20 & SEDclass & SED class (I, flat, II, III) \\
21 & dipper & LC matches dipper characteristics (see \S\ref{sec:LCPcats})\\
22 & burster & LC matches burster characteristics (see \S\ref{sec:LCPcats})\\
23 & single/multi-P &  single or multi-period star \\
24 & dd &  LC and power spectrum matches double-dip characteristics (see \S\ref{sec:LCPcats})\\
25 & ddmoving & LC and power spectrum matches moving double-dip characteristics (see \S\ref{sec:LCPcats})\\
26 & shapechanger & LC matches shape changer characteristics (see \S\ref{sec:LCPcats})\\
27 & beater &  LC has beating visible (see \S\ref{sec:LCPcats})\\
28 & complexpeak & power spectrum has a complex, structured peak and/or has a wide peak (see \S\ref{sec:LCPcats})\\
29 & resolvedclose & power spectrum has resolved close peaks (see \S\ref{sec:LCPcats})\\
30 & resolveddist & power spectrum has resolved distant peaks (see \S\ref{sec:LCPcats})\\
31 & pulsator & power spectrum and LC match pulsator characteristics (see \S\ref{sec:LCPcats})\\
\enddata
\end{deluxetable}

\section{Timescales}
\label{app:timescales}

As in papers I-V, some LCs have some repeated patterns that we cannot
identify with certainty as a rotation period. These `timescales' tend
to be longer than most of the rotation periods. Sometimes, there is
not enough data to go $>$1 complete cycle. Table~\ref{tab:timescales}
summarizes the timescales for the stars out of the entire ensemble.
Note that some also appear in the list of periodic stars, but with a
shorter period that we believe to be the rotation period; the
longer-term variability is unlikely to be rotation.

\floattable
\begin{deluxetable}{cccl}
\tabletypesize{\scriptsize}
\tablecaption{Lists of Objects with Timescales\label{tab:timescales}}
\tablewidth{0pt}
\tablehead{\colhead{EPIC} & \colhead{Other Name} 
&\colhead{Membership}&\colhead{Timescale} }
\startdata
210552148	&	LP415-113	&	NM	&	$\sim$35	\\
210666196	&	\ldots	&	NM	&	$\sim$15	\\
210683818	&	HBC393	&	Mem	&	$\sim$25	\\
210716897	&	RXJ0426.3+1836	&	NM	&
$\sim$38\tablenotemark{a}\\
246835564	&	\ldots	&	NM	&	$\sim$28	\\
247122185	&	\ldots	&	NM	&	$\sim$35	\\
247532735	&	V*V1108Tau	&	NM	&	$\sim$15\\	
247575425	&	IRAS04303+2240	&	Mem	&	$\sim$7.5\\	
247590222	&	IRAS04302+2247	&	Mem	&	$\sim$25\\	
247805410	&	V*FXTau	&	Mem	&	$\sim$11.6	\\
247820507	&	[RRA2004]Haro6-10VLA1	&	Mem	&	$\sim$13.6\\	
247946210	&	IRASF04570+2520	&	NM	&	$\sim$35	\\
247961728	&	V*V414Tau	&	NM	&	$\sim$35	\\
248019700	&	SSTtau044325.1+255706	&	NM	&	$\sim$23\\	
248049475	&	V*DOTau	&	Mem	&	$\sim$16	\\
248054085	&	SSTtau044453.1+261257	&	NM	&	$\sim$16.5\\	
\enddata
\tablenotetext{a}{This star has an additional real, shorter period,
which we take to be the rotation period.}
\end{deluxetable}

\section{Comparison to Literature Periods}
\label{app:lit}

In this appendix, we compare periods for stars in common between our
study and several literature studies, regardless of whether or not the
stars are currently thought to be Taurus members or not. Specific
stars are listed in Table~\ref{tab:comparelit}. 

\begin{figure}[h]
\epsscale{1.0}
\plotone{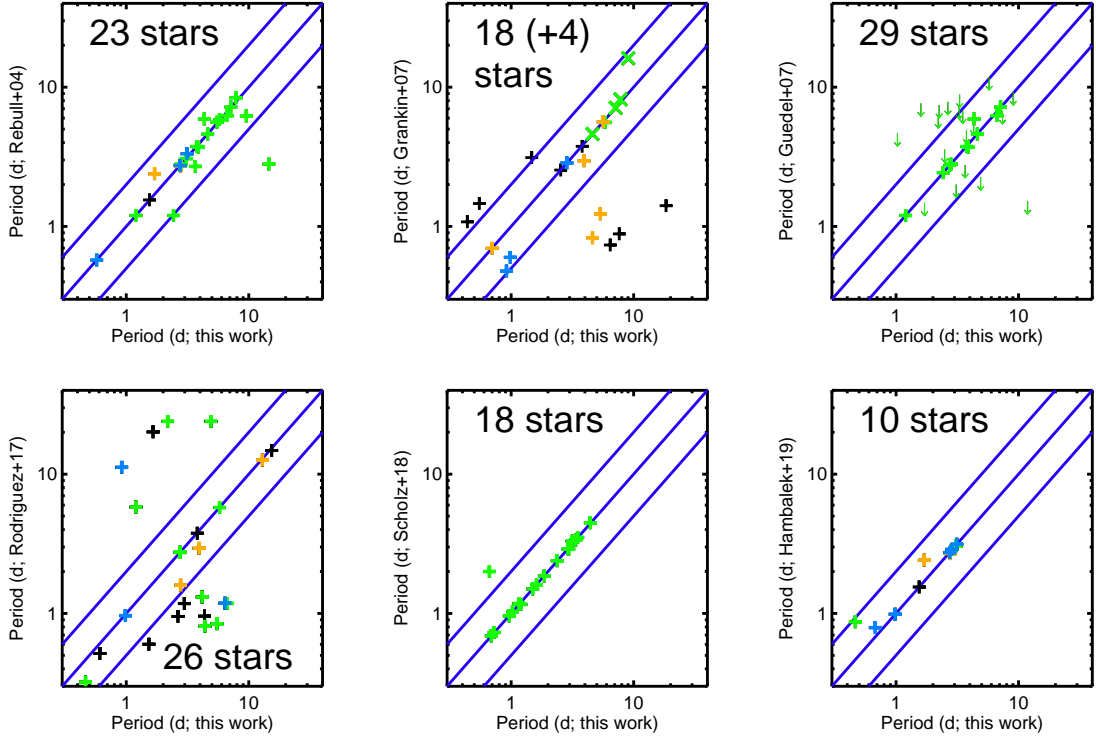}
\caption{Comparison of periods obtained here to periods obtained in
the literature.  Colors of points are: green= highest confidence
member; orange= possible member; blue=young, foreground object;
black=likely NM (see Sec.~\ref{sec:membership}. From upper
left to lower right: Rebull \etal\ (2004), literature compilation, 23
periodic stars in common; Grankin \etal\ (2007), 18 periodic stars in
common, with 4 more (shown as $\times$) from Artemenko \etal\ (2012);
G\"udel \etal\ (2007), literature compilation, 29 periodic stars in
common; Rodriguez \etal\ (2017b), KELT, 26 periodic stars in common;
Scholz \etal\ (2018), K2, 18 stars in common;  Hamb\'alek \etal\
(2019), SuperWASP/NSVS/K2, 10 stars in common. The dark blue lines are
at 1-to-1, $P$/2, and 2$P$.  (Xiao \etal\ 2012 has only 3 stars in
common, so is not plotted; see text and Table~\ref{tab:comparelit}.
Rigon \etal\ 2017 reports period ranges and so is not plotted here.)
Most of the periods match well (see the text for much more discussion). }
\label{fig:comparelit}
\end{figure}

Rebull \etal\ (2004) collected periods from the then-current
literature for many young stars, including stars in Taurus. There are
27 stars in common between the literature compilation in Rebull \etal\
(2004) and the present study, but only 23 of them are periodic in both
studies; see Figure~\ref{fig:comparelit}. Most (17) of them match
periods to within a fractional difference of 10\%, leaving 6 that
don't match. In three cases (EPIC 210690735=UX Tau), 247592463=HP Tau,
248029373=DK Tau), the LC is complex enough that it could explain the
different period obtained in the literature. For EPIC 210670948, the
period compiled in Rebull \etal\ (2004; citing Preibisch \& Smith
1997) does not match other literature nor the K2 period, and we
conclude it's wrong. For 246770655, the period from the literature
matches our secondary period (though the primary period appears in the
Figure).  In the last case, EPIC 247034775 (V1076 Tau), the literature
reports conflicting values. Rebull \etal\ (2004) recorded the period
as 6.2 d, apparently originally from Grankin (1993), and that is what
appears in Figure~\ref{fig:comparelit} and Table~\ref{tab:comparelit}.
Grankin (1997) reports that the period of 9.91 d from Grankin (1994)
``is more credible.'' The K2 data yield $P$=9.5412 d, supporting the
longer period; there is no $\sim$6 d period in the K2 data.

Grankin \etal\ (2007) presented periods for 39 stars in Taurus
obtained as part of the ROTOR program at Mount Maidanak Observatory.
Eighteen of those stars also have K2 LCs that we determine to be
periodic (4 more are not periodic in K2). Most (12) of these stars do
not have the same periods in the K2 data. Five of them are very
different periods, and five more are plausibly harmonics; see
Figure~\ref{fig:comparelit}. As part of this same ROTOR project,
Artemenko \etal\ (2012) published several additional periods. Of the 7
new stars in common, all of them have complex LCs. Only 4
have periods we derived from the K2 data; three of them match very
well, and the fourth may be a harmonic.

As part of the XMM-Newton Extended Survey of the Taurus Molecular
Cloud (XEST; G\"udel \etal\ 2007) and the Spitzer Taurus survey
(Rebull \etal\ 2010), an extensive survey of the literature at that
time was made, and a list of periods assembled to which we can 
compare. There are 40 stars in common with our study, only 29 of which
are periodic in both studies. However, most of those are curiously
listed as limits on the period in G\"udel \etal\ (2007); see
Figure~\ref{fig:comparelit}. For those stars with secure periods,
there is a good match, 8 of the 9 are well within 10\%. 

Xiao \etal\ (2012) used the Trans-atlantic Exoplanet Survey (TrES) to
monitor L1495, a portion of the Taurus molecular cloud. Only three
stars are in common between the studies: EPICs 247822311 (WK812),
248029373 (DK Tau), and 248040905 (IQ Tau). EPIC 248029373 is the only
one where the periods are not effectively identical. We obtain 7.84 d
where Xiao \etal\ report 4.094 d, which is not consistent with our data.
However, the K2 LC is complex and it is plausible that another
campaign at a different time may obtain a longer period.
 
Rigon \etal\ (2017) used WASP (the Wide Angle Search for Planets) to
probe the long-term variability of stars largely in Taurus-Auriga.
Only two stars are in common, AA Tau (EPIC 247810494; also see below)
and VO Tau (EPIC 248049475). Rigon \etal\ report period ranges, and we
do not find periods for either one. For the first one, they report a
58d and a 5d period; other literature report 8-13 d periods.  For the
latter, they report 5-6d, 24d, and 6-7d.  We see both of these stars
as irregular bursters during the K2 window, and cannot determine a
period; we would not be sensitive to periods longer than $\sim$35d in
any case.

Rodriguez \etal\ (2017b) used the Kilodegree Extremely Little
Telescope (KELT) to find rotation periods (and dippers) in Taurus.
Separately, the KELT LC for EPIC 248180268 = HD 283782 = V1334 Tau
appeared in Rodriguez \etal\ (2017a); since there is a period reported
there that came from KELT data, we include this period here.  In
total, there are 33 stars in common between us, 26 of which are
periodic in both studies. Only 7 match periods to within 10\%; see
Figure~\ref{fig:comparelit}. A few more have periods in KELT
comparable to our secondary or teriary period; see
Table~\ref{tab:comparelit}. For each of the stars in common, the LCs
were investigated in detail, and the KELT period could not be
recovered in most cases. We find a few stars appear as irregular
bursters (at least during the K2 epoch), so perhaps it is not
surprising in those cases that the periods are so different. There
is no clear reason why the rest of the stars are not recovered at the
same periods. However, many of the KELT LCs that do not match are of
very low amplitude, and the KELT LCs cover a very long time baseline,
so if, say, periods are appearing/disappearing at different latitudes
under the assumption of surface differential rotation, the rotation
period may appear to change over the KELT baseline, making it harder
to determine. For the LCs we have, our periods are correct, though
some of the LCs are complex. (We note that our recovery rate between
K2 and KELT was much higher in our earlier K2 papers.)

More recently, Scholz \etal\ (2018) used some of the same K2 data as
we are using to explore rotation rates of brown dwarfs in Taurus, and
there are 18 stars in common. The periods are all recovered well,
except EPIC 248023915 (2MASSJ04380083+2558572), where the period
reported in Scholz \etal\ is about twice the second period we obtain;
we suspect they found an alias rather than the true period. 

Most recently, Hamb\'alek \etal\ (2019) used SuperWASP, the Northern
Sky Variability Survey (NSVS), and K2 to study 20 weak-lined T~Tauri
stars in Taurus; there are 12 stars in common, 10 of which are
periodic in both studies. Most of the stars are well matched; see
Figure~\ref{fig:comparelit}. There are three stars where we report two
K2 periods, and Hamb\'alek \etal\ find the period we identified as the
second period: EPICs 247076294, 248180268, and 246770655. In the case
of 247076294, the two K2 periods are so close to each other that the
difference does not stand out in the plot, but the Hamb\'alek \etal\
period matches the second K2 period.  

AA~Tau (EPIC 247810494) is somewhat of a special case. It has been the
subject of numerous, intensive studies (e.g., Donati \etal\ 2010;
Bouvier \etal\ 2007, 1999; and references therein). Those studies
identified AA~Tau as having a period of 8.22d and defining the class
of ``dippers'' among disk-bearing stars regularly occulted by disk
matter close to the star. There is no evidence for an $\sim$8 d period
in the K2 data; in fact, we find no significant periodicity at all.
Additionally, we find no obvious dipper signatures during this
campaign; particularly in the first half of the campaign, the LC
shares more characteristics with bursters than dippers. Substantial
variation has been noted in this system before (e.g., Bouvier \etal\
2003, 2013).  The dipper character of the light curve as well as the
8.2 d period seen in ground-based light curves for decades disappeared
altogether as the star  underwent a major dimming event in recent
years (Bouvier \etal\ 2013), a state from which it has not yet
recovered.

\clearpage

\startlongtable
\floattable
\begin{longrotatetable}
\begin{deluxetable}{cp{2cm}cccccccccccccp{5cm}}
\tabletypesize{\scriptsize}
\floattable
\centerwidetable
\tablecaption{Comparison to Literature Periods (all in units of days)\label{tab:comparelit}}
\tablewidth{0pt}
\tablehead{\colhead{EPIC} & \colhead{Other Name}
& \colhead{Rebull} & \colhead{Grankin} & \colhead{Artemenko}& \colhead{G\"udel} 
& \colhead{Xiao} & \colhead{Rodriguez} 
& \colhead{Scholz} & \colhead{Hamb\'alek} 
& \colhead{$P_1$} & \colhead{$P_2$} &
\colhead{$P_3$} & \colhead{$P_4$} &
\colhead{Match?} & \colhead{Notes}\\ [-0.4cm]
& 
& \colhead{+04} & \colhead{+07 } & \colhead{+12 }& \colhead{+07 } 
& \colhead{+17} & \colhead{+17b } 
& \colhead{+18} & \colhead{+19} 
}
\startdata
210631263	&	 HD285372 	&	0.573	&	\ldots	&	\ldots	&	  	\ldots	&	\ldots	&	\ldots	&	\ldots	&	\ldots	&	0.5733	&	\ldots	&	\ldots	&	\ldots	&	y	&	\ldots	\\
210662824	&	 HD285778 	&	2.74	&	\ldots	&	\ldots	&	  	\ldots	&	\ldots	&	\ldots	&	\ldots	&	2.736	&	2.7503	&	1.3713	&	\ldots	&	\ldots	&	y	&	\ldots	\\
210670948	&	 [FK83]LDN155151 	&	1.2	&	\ldots	&	\ldots	&	 	2.43	&	\ldots	&	\ldots	&	\ldots	&	\ldots	&	2.4282	&	\ldots	&	\ldots	&	\ldots	&	y/n	&	Match with G\"udel+ but not Rebull+.	\\
210674635	&	 2MASS J04312405+1800215 	&	\ldots	&	\ldots	&	\ldots	&	 $<$	6.29	&	\ldots	&	\ldots	&	\ldots	&	\ldots	&	2.2098	&	\ldots	&	\ldots	&	\ldots	&	n	&	True that $P_{\rm here}$ is $<P_{\rm XEST}$, but 2.2d really clear in K2, and nothing at $\sim$6 d.	\\
210683818	&	 HBC393 	&	\ldots	&	\ldots	&	\ldots	&	 $<$	3.82	&	\ldots	&	\ldots	&	\ldots	&	\ldots	&	\ldots	&	\ldots	&	\ldots	&	\ldots	&	n	&	At best, K2 data suggest a 20-25d timescale; there isn't anything near 3.8 d.	\\
210689130	&	 HD28150 	&	\ldots	&	0.696	&	\ldots	&	  	\ldots	&	\ldots	&	\ldots	&	\ldots	&	\ldots	&	0.6981	&	\ldots	&	\ldots	&	\ldots	&	y	&	\ldots	\\
210690598	&	 EM*LkHA358 	&	\ldots	&	\ldots	&	\ldots	&	 $<$	6.8	&	\ldots	&	\ldots	&	\ldots	&	\ldots	&	3.5045	&	3.7378	&	\ldots	&	\ldots	&	n	&	True that $P_{\rm here}$ is $<P_{\rm XEST}$, but $\sim$3d really clear in K2, and nothing at $\sim$7 d.	\\
210690735	&	 V*UXTau 	&	2.7	&	\ldots	&	\ldots	&	  	\ldots	&	\ldots	&	\ldots	&	\ldots	&	\ldots	&	3.6584	&	\ldots	&	\ldots	&	\ldots	&	n	&	No match with Rebull+ from either period. However, K2 LC power spectrum suggests $\sim$2 d at another time may be plausible.	\\
210690892	&	 XZTauAB 	&	2.6	&	\ldots	&	3.24	&	 	2.6	&	\ldots	&	\ldots	&	\ldots	&	\ldots	&	\ldots	&	\ldots	&	\ldots	&	\ldots	&	n	&	No period in K2.	\\
210690913	&	 V*HLTau 	&	\ldots	&	\ldots	&	\ldots	&	 $<$	7.21	&	\ldots	&	\ldots	&	\ldots	&	\ldots	&	\ldots	&	\ldots	&	\ldots	&	\ldots	&	n	&	There is a 7.7d peak in the K2 data but it does not phase well; a 24d period is more plausible in the K2 data.	\\
210698143	&	 2MASS J04311578+1820072 	&	\ldots	&	\ldots	&	\ldots	&	 $<$	7.66	&	\ldots	&	\ldots	&	\ldots	&	\ldots	&	1.5954	&	3.4961	&	\ldots	&	\ldots	&	n	&	True that $P_{\rm here}$ is $<P_{\rm XEST}$, but periods really clear in K2, and nothing at $\sim$7 d.	\\
210698281	&	 V*V827Tau 	&	3.75	&	\ldots	&	\ldots	&	 	3.75	&	\ldots	&	\ldots	&	\ldots	&	\ldots	&	3.7582	&	\ldots	&	\ldots	&	\ldots	&	y	&	\ldots	\\
210699801	&	 EM*LkHA267 	&	\ldots	&	\ldots	&	\ldots	&	 $<$	5.67	&	\ldots	&	\ldots	&	\ldots	&	\ldots	&	4.3047	&	1.07	&	\ldots	&	\ldots	&	n	&	True that $P_{\rm here}$ is $<P_{\rm XEST}$, but periods clear in K2 and nothing at $\sim$5.6 d.	\\
210767482	&	 HD284418 	&	\ldots	&	\ldots	&	\ldots	&	  	\ldots	&	\ldots	&	14.81	&	\ldots	&	\ldots	&	15.3682	&	\ldots	&	\ldots	&	\ldots	&	y	&	$\sim$match	\\
210777988	&	 V*TTau 	&	2.8	&	\ldots	&	\ldots	&	 	2.8	&	\ldots	&	\ldots	&	\ldots	&	\ldots	&	2.8116	&	\ldots	&	\ldots	&	\ldots	&	y	&	\ldots	\\
210792668	&	\ldots	&	\ldots	&	1.168	&	\ldots	&	  	\ldots	&	\ldots	&	\ldots	&	\ldots	&	\ldots	&	\ldots	&	\ldots	&	\ldots	&	\ldots	&	n	&	No period in K2.	\\
210805120	&	 V*V1196Tau 	&	3.02	&	\ldots	&	\ldots	&	  	\ldots	&	\ldots	&	\ldots	&	\ldots	&	\ldots	&	\ldots	&	\ldots	&	\ldots	&	\ldots	&	y	&	No period in K2.	\\
210818897	&	 V*V1298Tau 	&	\ldots	&	2.86	&	\ldots	&	  	\ldots	&	\ldots	&	\ldots	&	\ldots	&	2.885	&	2.8484	&	1.4508	&	\ldots	&	\ldots	&	y	&	\ldots	\\
210881343	&	  	&	\ldots	&	0.736	&	\ldots	&	  	\ldots	&	\ldots	&	\ldots	&	\ldots	&	\ldots	&	6.4388	&	2.7789	&	\ldots	&	\ldots	&	n	&	No period near 0.7 in K2 data.	\\
210977750	&	 2MASS J04073502+2237394 	&	\ldots	&	0.816	&	\ldots	&	  	\ldots	&	\ldots	&	\ldots	&	\ldots	&	\ldots	&	0.1346	&	0.1507	&	\ldots	&	\ldots	&	n	&	No period near 0.8 in K2 data.	\\
211104793	&	 2MASS J04124068+2438157 	&	\ldots	&	5.58	&	\ldots	&	  	\ldots	&	\ldots	&	\ldots	&	\ldots	&	\ldots	&	5.8543	&	\ldots	&	\ldots	&	\ldots	&	y	&	\ldots	\\
211182143	&	\ldots	&	\ldots	&	1.961	&	\ldots	&	  	\ldots	&	\ldots	&	\ldots	&	\ldots	&	\ldots	&	\ldots	&	\ldots	&	\ldots	&	\ldots	&	n	&	No period in K2.	\\
246695532	&	 V*V1840Ori 	&	\ldots	&	1.231	&	\ldots	&	  	\ldots	&	\ldots	&	\ldots	&	\ldots	&	\ldots	&	5.343	&	\ldots	&	\ldots	&	\ldots	&	n	&	There is a peak near 1.5 d in one and only one of the data reductions (K2SFF); we suspect that this period may correspond to one of the other targets near 246695532.	\\
246760205	&	 HD286179 	&	3.33	&	\ldots	&	\ldots	&	  	\ldots	&	\ldots	&	\ldots	&	\ldots	&	3.14	&	3.1297	&	\ldots	&	\ldots	&	\ldots	&	y	&	\ldots	\\
246770655	&	 WDS J04573+1524AB 	&	2.39	&	\ldots	&	\ldots	&	  	\ldots	&	\ldots	&	\ldots	&	\ldots	&	2.412	&	1.7013	&	2.3551	&	\ldots	&	\ldots	&	y	&	$\sim$match with our $P_2$	\\
246798563	&	 V*V1326Tau 	&	\ldots	&	2.54	&	\ldots	&	  	\ldots	&	\ldots	&	\ldots	&	\ldots	&	\ldots	&	2.5385	&	\ldots	&	\ldots	&	\ldots	&	y	&	\ldots	\\
246802680	&	 HD285957 	&	3.07	&	\ldots	&	\ldots	&	  	\ldots	&	\ldots	&	\ldots	&	\ldots	&	3.055	&	3.0868	&	\ldots	&	\ldots	&	\ldots	&	y	&	\ldots	\\
246815623	&	 V*V1351Tau 	&	\ldots	&	5.64	&	\ldots	&	  	\ldots	&	\ldots	&	\ldots	&	\ldots	&	\ldots	&	5.6313	&	\ldots	&	\ldots	&	\ldots	&	y	&	\ldots	\\
246825172	&	 V*V1353Tau 	&	\ldots	&	0.884	&	\ldots	&	  	\ldots	&	\ldots	&	\ldots	&	\ldots	&	\ldots	&	7.6307	&	\ldots	&	\ldots	&	\ldots	&	n	&	No period near 0.9 d in K2 data.	\\
246923113	&	 V*DRTau 	&	2.8	&	\ldots	&	\ldots	&	  	\ldots	&	\ldots	&	\ldots	&	\ldots	&	\ldots	&	14.5698	&	\ldots	&	\ldots	&	\ldots	&	n	& 	No period near 2.8 d in K2 data.	\\
246977631	&	 V*V1346Tau 	&	\ldots	&	0.829	&	\ldots	&	  	\ldots	&	\ldots	&	\ldots	&	\ldots	&	\ldots	&	4.6167	&	\ldots	&	\ldots	&	\ldots	&	n	&	No period near 0.8 d in K2 data.	\\
247031423	&	 HD28867 	&	\ldots	&	\ldots	&	\ldots	&	 $<$	2	&	\ldots	&	\ldots	&	\ldots	&	\ldots	&	3.0978	&	3.2775	&	\ldots	&	\ldots	&	n	&	No match even with the inequality; perhaps G\"uedel+ found a harmonic?	\\
247032616	&	 V*V826Tau 	&	3.7	&	\ldots	&	\ldots	&	 	3.7	&	\ldots	&	\ldots	&	\ldots	&	\ldots	&	3.8829	&	\ldots	&	\ldots	&	\ldots	&	y	&	\ldots	\\
247034775	&	 V*V1076Tau 	&	6.2	&	\ldots
&	\ldots	&	  	\ldots	&	\ldots	&	\ldots
&	\ldots	&	\ldots	&	9.5412	&	\ldots	& \ldots	&	\ldots	&	n	&	No period near 6 d in K2 data, but 6d may not be real; 9.91 d from Grankin (1994) is a better match to the K2 data.	\\
247047380	&	 V*DMTau 	&	\ldots	&	\ldots	&	\ldots	&	 $<$	6.74	&	\ldots	&	\ldots	&	\ldots	&	\ldots	&	7.4081	&	\ldots	&	\ldots	&	\ldots	&	n	&	Not a match, but the K2 LC has potentially many periods in it, including near 6 d.	\\
247051861	&	 2MASS J04321606+1812464 	&	\ldots	&	\ldots	&	\ldots	&	 $<$	7.84	&	\ldots	&	\ldots	&	\ldots	&	\ldots	&	2.6496	&	\ldots	&	\ldots	&	\ldots	&	n	&	True that $P_{\rm here}$ is $<P_{\rm XEST}$, but $\sim$3 d really clear in K2, and nothing at $\sim$8 d.	\\
247076294	&	 HD31281 	&	\ldots	&	\ldots	&	\ldots	&	  	\ldots	&	\ldots	&	\ldots	&	\ldots	&	0.791	&	0.6771	&	0.7999	&	\ldots	&	\ldots	&	y	&	Hambalek+ found our $P_2$.	\\
247119725	&	 [WKS96]42 	&	\ldots	&	\ldots	&	\ldots	&	  	\ldots	&	\ldots	&	1.31	&	\ldots	&	\ldots	&	4.175	&	\ldots	&	\ldots	&	\ldots	&	n	&	No period near 1.3 d in K2 data.	\\
247126197	&	 HD285840 	&	1.55	&	\ldots	&	\ldots	&	  	\ldots	&	\ldots	&	\ldots	&	\ldots	&	1.548	&	1.5461	&	\ldots	&	\ldots	&	\ldots	&	y	&	\ldots	\\
247139505	&	 LP415-183 	&	\ldots	&	1.41	&	\ldots	&	  	\ldots	&	\ldots	&	\ldots	&	\ldots	&	\ldots	&	18.3841	&	\ldots	&	\ldots	&	\ldots	&	n	&	No period near 1.4 d in K2 data.	\\
247225984	&	 V*V1333Tau 	&	\ldots	&	\ldots	&	\ldots	&	  	\ldots	&	\ldots	&	12.594	&	\ldots	&	\ldots	&	12.9213	&	\ldots	&	\ldots	&	\ldots	&	y	&	\ldots	\\
247280905	&	 V*V1325Tau 	&	\ldots	&	2.96	&	\ldots	&	  	\ldots	&	\ldots	&	2.935	&	\ldots	&	\ldots	&	3.9332	&	2.9233	&	11.7995	&	\ldots	&	y	&	Others match our $P_2$.	\\
247303990	&	 2MASS J04564525+2035116 	&	\ldots	&	1.46	&	\ldots	&	  	\ldots	&	\ldots	&	\ldots	&	\ldots	&	\ldots	&	0.5498	&	0.793	&	\ldots	&	\ldots	&	n	&	Complicated K2 LC, with several strong peaks, but none of them are at $\sim$1.5 d, and none of them are harmonics of 1.5 d.	\\
247360583	&	 2MASS J05064662+2104296 	&	\ldots	&	1.079	&	\ldots	&	  	\ldots	&	\ldots	&	\ldots	&	\ldots	&	\ldots	&	0.4389	&	\ldots	&	\ldots	&	\ldots	&	n	&	No period near 1.1 d in K2 data.	\\
247454835	&	 HD284496 	&	2.71	&	\ldots	&	\ldots	&	  	\ldots	&	\ldots	&	1.593	&	\ldots	&	2.688	&	2.7733	&	2.653	&	\ldots	&	\ldots	&	y/n	&	Rodriguez+ is the outlier.	\\
247520207	&	 EM*LkCa15 	&	5.85	&	\ldots	&	\ldots	&	  	\ldots	&	\ldots	&	5.765	&	\ldots	&	\ldots	&	5.7795	&	\ldots	&	\ldots	&	\ldots	&	y	&	\ldots	\\
247528573	&	 2MASS J04464475+2224508 	&	\ldots	&	\ldots	&	\ldots	&	  	\ldots	&	\ldots	&	0.517	&	\ldots	&	\ldots	&	\ldots	&	\ldots	&	\ldots	&	\ldots	&	n	&	No period in K2.	\\
247539775	&	 V*V1341Tau 	&	\ldots	&	0.478	&	\ldots	&	  	\ldots	&	\ldots	&	11.198	&	\ldots	&	\ldots	&	0.9165	&	\ldots	&	\ldots	&	\ldots	&	n	&	Very, very clear period in K2, not near 0.5 or 11 d.	\\
247548866	&	 [BLH2002]KPNO-Tau8 	&	\ldots	&	\ldots	&	\ldots	&	  	\ldots	&	\ldots	&	\ldots	&	0.69	&	\ldots	&	0.6878	&	\ldots	&	\ldots	&	\ldots	&	y	&	\ldots	\\
247575425	&	 IRAS04303+2240 	&	\ldots	&	\ldots	&	\ldots	&	 $<$	5.23	&	\ldots	&	\ldots	&	\ldots	&	\ldots	&	\ldots	&	\ldots	&	\ldots	&	\ldots	&	n	&	Complicated K2 LC power spectrum suggests $\sim$5.x d at another time may be plausible.	\\
247575958	&	 2MASS J04330945+2246487 	&	\ldots	&	\ldots	&	\ldots	&	  	\ldots	&	\ldots	&	\ldots	&	3.51	&	\ldots	&	3.4916	&	\ldots	&	\ldots	&	\ldots	&	y	&	\ldots	\\
247584113	&	 CITau 	&	\ldots	&	\ldots	&	16.1	&	 $<$	9.2	&	\ldots	&	\ldots	&	\ldots	&	\ldots	&	9.0335	&	6.6224	&	\ldots	&	\ldots	&	n	&	   More $P$ in literature are 8.9891,16.10 (Johns Krull \etal\ 2016). $\sim$9 d is thought to be period of planet, which is recovered; nothing near 16d in K2, but LC complicated. More $P$ in literature (based on this K2 LC) is $\sim$6.6 and $\sim$9, for star and planet, respectively (Biddle \etal\ 2018).	\\
247585953	&	 2MASS J04340717+2251227 	&	\ldots	&	\ldots	&	\ldots	&	  	\ldots	&	\ldots	&	1.177	&	\ldots	&	\ldots	&	2.9775	&	5.3517	&	1.4534	&	\ldots	&	n	&	No period near 1.7 d in K2 data.	\\
247591534	&	 2MASS J04355760+2253574 	&	\ldots	&	\ldots	&	\ldots	&	  	\ldots	&	\ldots	&	\ldots	&	1.16	&	\ldots	&	1.1984	&	\ldots	&	\ldots	&	\ldots	&	y	&	\ldots	\\
247592103	&	 CoKuHPTauG2 	&	1.2	&	\ldots	&	\ldots	&	 	1.2	&	\ldots	&	5.798	&	\ldots	&	\ldots	&	1.1978	&	1.222	&	\ldots	&	\ldots	&	y/n	&	Rodriguez+ is the outlier.	\\
247592463	&	 V*HPTau 	&	5.9	&	\ldots	&	\ldots	&	 	5.9	&	\ldots	&	\ldots	&	\ldots	&	\ldots	&	4.3307	&	\ldots	&	\ldots	&	\ldots	&	n	&	Complex K2 LC, could explain mismatch. 	\\
247592919	&	 Haro6-28 	&	\ldots	&	\ldots	&	\ldots	&	 $<$	3.98	&	\ldots	&	\ldots	&	\ldots	&	\ldots	&	\ldots	&	\ldots	&	\ldots	&	\ldots	&	n	&	Complex K2 LC, no clear periods.	\\
247594260	&	 V*V1335Tau 	&	\ldots	&	3.762	&	\ldots	&	  	\ldots	&	\ldots	&	3.766	&	\ldots	&	\ldots	&	3.7998	&	\ldots	&	\ldots	&	\ldots	&	y	&	\ldots	\\
247599080	&	 IRAS04295+2251 	&	\ldots	&	\ldots	&	\ldots	&	 $<$	1.48	&	\ldots	&	\ldots	&	\ldots	&	\ldots	&	1.7081	&	\ldots	&	\ldots	&	\ldots	&	n	&	No match even with the inequality; K2 LC is low SNR.	\\
247600777	&	 2MASS J04363893+2258119 	&	\ldots	&	\ldots	&	\ldots	&	  	\ldots	&	\ldots	&	\ldots	&	0.96	&	\ldots	&	0.9645	&	\ldots	&	\ldots	&	\ldots	&	y	&	\ldots	\\
247604448	&	 2MASS J04361038+2259560 	&	\ldots	&	\ldots	&	\ldots	&	  	\ldots	&	\ldots	&	\ldots	&	2.91	&	\ldots	&	2.933	&	\ldots	&	\ldots	&	\ldots	&	y	&	\ldots	\\
247609913	&	 V*V1117Tau 	&	\ldots	&	\ldots	&	\ldots	&	  	\ldots	&	\ldots	&	1.184	&	\ldots	&	\ldots	&	6.3887	&	\ldots	&	\ldots	&	\ldots	&	n	&	No period near 1.2 d in K2 data.	\\
247630187	&	 2MASS J04350850+2311398 	&	\ldots	&	\ldots	&	\ldots	&	  	\ldots	&	\ldots	&	\ldots	&	1.5	&	\ldots	&	1.4981	&	\ldots	&	\ldots	&	\ldots	&	y	&	\ldots	\\
247739445	&	 2MASS J04302365+2359129 	&	\ldots	&	\ldots	&	\ldots	&	  	\ldots	&	\ldots	&	\ldots	&	1.61	&	\ldots	&	1.6119	&	\ldots	&	\ldots	&	\ldots	&	y	&	\ldots	\\
247748412	&	 2MASS J04322329+2403013 	&	\ldots	&	\ldots	&	\ldots	&	  	\ldots	&	\ldots	&	\ldots	&	3.37	&	\ldots	&	3.3643	&	\ldots	&	\ldots	&	\ldots	&	y	&	\ldots	\\
247763883	&	 V*GHTau 	&	\ldots	&	\ldots	&	\ldots	&	 $<$	3.57	&	\ldots	&	\ldots	&	\ldots	&	\ldots	&	2.4937	&	2.9408	&	\ldots	&	\ldots	&	n	&	True that $P_{\rm here}$ is $<P_{\rm XEST}$; K2 LC power spectrum suggests a longer period at another time may be plausible. 	\\
247764745	&	 V*V807Tau 	&	\ldots	&	\ldots	&	\ldots	&	  	\ldots	&	\ldots	&	0.809	&	\ldots	&	\ldots	&	4.3784	&	\ldots	&	\ldots	&	\ldots	&	n	&	No period near 0.8 d in K2 data.	\\
247776236	&	 HD28975 	&	\ldots	&	\ldots	&	\ldots	&	  	\ldots	&	\ldots	&	0.957	&	\ldots	&	\ldots	&	4.353	&	\ldots	&	\ldots	&	\ldots	&	n	&	No period near 0.9 d in K2 data.	\\
247781229	&	 HD28819 	&	\ldots	&	\ldots	&	\ldots	&	  	\ldots	&	\ldots	&	0.516	&	\ldots	&	\ldots	&	0.6086	&	0.2849	&	\ldots	&	\ldots	&	n	&	K2 LC power spectrum suggests that $\sim$0.5 d at another time may be plausible.	\\
247791556	&	 2MASS J04330197+2421000 	&	\ldots	&	\ldots	&	\ldots	&	 $<$	4.67	&	\ldots	&	\ldots	&	1.03	&	\ldots	&	1.0261	&	1.0469	&	\ldots	&	\ldots	&	n	&	True that $P_{\rm here}$ is $<P_{\rm XEST}$; nothing near 4.6 d in K2.	\\
247791801	&	 V*GKTau 	&	4.6	&	\ldots	&	4.61	&	 	4.6	&	\ldots	&	\ldots	&	\ldots	&	\ldots	&	4.6165	&	\ldots	&	\ldots	&	\ldots	&	y	&	\ldots	\\
247792225	&	 V*GITau 	&	7.2	&	\ldots	&	7.09	&	 	7.2	&	\ldots	&	\ldots	&	\ldots	&	\ldots	&	7.1334	&	\ldots	&	\ldots	&	\ldots	&	y	&	\ldots	\\
247795097	&	 V*V928Tau 	&	\ldots	&	\ldots	&	\ldots	&	 $<$	7.45	&	\ldots	&	\ldots	&	\ldots	&	\ldots	&	2.2468	&	2.4876	&	\ldots	&	\ldots	&	n	&	True that $P_{\rm here}$ is $<P_{\rm XEST}$; K2 periods very clear and nothing near 7 d.	\\
247799571	&	 V*HKTau 	&	\ldots	&	\ldots	&	\ldots	&	 $<$	8.86	&	\ldots	&	\ldots	&	\ldots	&	\ldots	&	3.3056	&	\ldots	&	\ldots	&	\ldots	&	n	&	True that $P_{\rm here}$ is $<P_{\rm XEST}$; K2 LC power spectrum suggests a longer period at another time may be plausible. 	\\
247804500	&	 2MASS J04352474+2426218 	&	\ldots	&	\ldots	&	\ldots	&	  	\ldots	&	\ldots	&	0.929	&	\ldots	&	\ldots	&	\ldots	&	\ldots	&	\ldots	&	\ldots	&	n	&	No period in K2.	\\
247805410	&	 V*FXTau 	&	\ldots	&	\ldots	&	\ldots	&	 $<$	12.34	&	\ldots	&	\ldots	&	\ldots	&	\ldots	&	\ldots	&	\ldots	&	\ldots	&	\ldots	&	n	&	Complicated K2 LC power spectrum suggests a longer period at another time may be plausible. 	\\
247810494	&	 V*AATau 	&	8.22	&	\ldots	&	8.19	&	 	8.22	&	\ldots	&	13.613	&	\ldots	&	\ldots	&	\ldots	&	\ldots	&	\ldots	&	\ldots	&	y/n	&	Rigon \etal\ (2017) reports  58d, 5d. No period in K2, but LC has a lot of structure, suggesting that periodic behavior at another time may be plausible.	\\
247810751	&	 Haro6-13 	&	\ldots	&	\ldots	&	\ldots	&	 $<$	7.3	&	\ldots	&	\ldots	&	\ldots	&	\ldots	&	3.2778	&	\ldots	&	\ldots	&	\ldots	&	n	&	True that $P_{\rm here}$ is $<P_{\rm XEST}$; K2 LC power spectrum suggests a longer period at another time may be plausible. 	\\
247820507	&	 [RRA2004]Haro6-10VLA1 	&	\ldots	&	\ldots	&	\ldots	&	 $<$	5.63	&	\ldots	&	\ldots	&	\ldots	&	\ldots	&	\ldots	&	\ldots	&	\ldots	&	\ldots	&	n	&	No match even with the inequality; K2 LC power spectrum suggests a shorter period at another time may be plausible. 	\\
247822311	&	 WK812 	&	2.75	&	\ldots	&	\ldots	&	 	2.75	&	2.747	&	2.743	&	\ldots	&	\ldots	&	2.7412	&	\ldots	&	\ldots	&	\ldots	&	y	&	\ldots	\\
247837468	&	 IRAS04264+2433 	&	\ldots	&	\ldots	&	\ldots	&	 $<$	1.52	&	\ldots	&	\ldots	&	\ldots	&	\ldots	&	11.8717	&	\ldots	&	\ldots	&	\ldots	&	n	&	No match even with the inequality; K2 LC power spectrum suggests a shorter period at another time may be plausible. 	\\
247843485	&	 V*ZZTau 	&	\ldots	&	\ldots	&	\ldots	&	  	\ldots	&	\ldots	&	1.311	&	\ldots	&	\ldots	&	4.1609	&	\ldots	&	\ldots	&	\ldots	&	n	&	No period near 1.3 d in K2 data.	\\
247864498	&	 V*IWTau 	&	5.6	&	\ldots	&	\ldots	&	  	\ldots	&	\ldots	&	0.844	&	\ldots	&	\ldots	&	5.5013	&	7.0425	&	\ldots	&	\ldots	&	y/n	&	Rodriguez+ is the outlier.	\\
247915927	&	 IRASS04414+2506 	&	\ldots	&	\ldots	&	\ldots	&	  	\ldots	&	\ldots	&	\ldots	&	4.48	&	\ldots	&	4.43	&	\ldots	&	\ldots	&	\ldots	&	y	&	\ldots	\\
247923794	&	 V*DPTau 	&	\ldots	&	\ldots	&	\ldots	&	 $<$	2.76	&	\ldots	&	\ldots	&	\ldots	&	\ldots	&	3.6623	&	\ldots	&	\ldots	&	\ldots	&	n	&	No match even with the inequality; K2 LC power spectrum suggests a shorter period at another time may be plausible. 	\\
247935061	&	 V*GOTau 	&	\ldots	&	\ldots	&	\ldots	&	 $<$	3.96	&	\ldots	&	\ldots	&	\ldots	&	\ldots	&	\ldots	&	\ldots	&	\ldots	&	\ldots	&	n	&	No period in K2, but LC has a lot of structure, suggesting that periodic behavior at another time may be plausible.	\\
247941378	&	 V*V999Tau 	&	\ldots	&	\ldots	&	\ldots	&	 $<$	2.25	&	\ldots	&	23.935	&	\ldots	&	\ldots	&	4.9052	&	2.6515	&	\ldots	&	\ldots	&	n	&	Two K2 periods are very clear, neither consistent with prior observations.	\\
247941613	&	 V*V1000Tau 	&	\ldots	&	\ldots	&	\ldots	&	  	\ldots	&	\ldots	&	23.935	&	\ldots	&	\ldots	&	2.1764	&	8.2537	&	\ldots	&	\ldots	&	n	&	Complex K2 LC, could explain mismatch. 	\\
247941930	&	 EM*LkHA332 	&	\ldots	&	\ldots	&	\ldots	&	 $<$	4.9	&	\ldots	&	23.935	&	\ldots	&	\ldots	&	\ldots	&	\ldots	&	\ldots	&	\ldots	&	n	&	Complex K2 LC, could be plausible at another time. 	\\
247950452	&	 2MASS J04334291+2526470 	&	\ldots	&	\ldots	&	\ldots	&	  	\ldots	&	\ldots	&	\ldots	&	0.73	&	\ldots	&	0.7269	&	\ldots	&	\ldots	&	\ldots	&	y	&	\ldots	\\
247953586	&	 2MASS J04320329+2528078 	&	\ldots	&	\ldots	&	\ldots	&	  	\ldots	&	\ldots	&	\ldots	&	2.39	&	\ldots	&	2.3765	&	\ldots	&	\ldots	&	\ldots	&	y	&	\ldots	\\
247968420	&	 2MASS J04414825+2534304 	&	\ldots	&	\ldots	&	\ldots	&	  	\ldots	&	\ldots	&	\ldots	&	2.9	&	\ldots	&	2.9144	&	\ldots	&	\ldots	&	\ldots	&	y	&	\ldots	\\
247986526	&	 DFTauAB 	&	7.2	&	\ldots	&	7.18	&	  	\ldots	&	\ldots	&	16.504	&	\ldots	&	\ldots	&	\ldots	&	\ldots	&	\ldots	&	\ldots	&	n	&	Complex K2 LC, could be plausible at another time. 	\\
247991214	&	 2MASS J04390396+2544264 	&	\ldots	&	\ldots	&	\ldots	&	  	\ldots	&	\ldots	&	\ldots	&	3.3	&	\ldots	&	3.1401	&	\ldots	&	\ldots	&	\ldots	&	y	&	\ldots	\\
247992574	&	 V*GNTau 	&	\ldots	&	\ldots	&	\ldots	&	 $<$	11.8	&	\ldots	&	\ldots	&	\ldots	&	\ldots	&	5.7452	&	\ldots	&	\ldots	&	\ldots	&	n	&	True that $P_{\rm here}$ is $<P_{\rm XEST}$; LC has a lot of structure, suggesting that a different period at another time may be plausible.	\\
248009353	&	 UZTauAB 	&	\ldots	&	\ldots	&	\ldots	&	 $<$	4.9	&	\ldots	&	\ldots	&	\ldots	&	\ldots	&	3.7527	&	\ldots	&	\ldots	&	\ldots	&	n	&	True that $P_{\rm here}$ is $<P_{\rm XEST}$; LC has a lot of structure, suggesting that a different period at another time may be plausible.	\\
248010721	&	 HD283707 	&	\ldots	&	\ldots	&	\ldots	&	  	\ldots	&	\ldots	&	0.967	&	\ldots	&	\ldots	&	\ldots	&	\ldots	&	\ldots	&	\ldots	&	n	&	No period in K2.	\\
248014510	&	 HD283809 	&	\ldots	&	\ldots	&	\ldots	&	  	\ldots	&	\ldots	&	20.121	&	\ldots	&	\ldots	&	1.653	&	1.4033	&	0.9651	&	2.3834	&	n	&	No period near 20 d in K2 data.	\\
248018164	&	 Haro6-33 	&	\ldots	&	\ldots	&	\ldots	&	 $<$	2.13	&	\ldots	&	\ldots	&	\ldots	&	\ldots	&	\ldots	&	\ldots	&	\ldots	&	\ldots	&	n	&	No period in K2, but LC has a lot of structure, suggesting that periodic behavior at another time may be plausible.	\\
248018652	&	 [BLH2002]KPNO-Tau7 	&	\ldots	&	\ldots	&	\ldots	&	  	\ldots	&	\ldots	&	\ldots	&	1.18	&	\ldots	&	1.1572	&	\ldots	&	\ldots	&	\ldots	&	y	&	\ldots	\\
248019693	&	 XEST02-040 	&	\ldots	&	3.12	&	\ldots	&	  	\ldots	&	\ldots	&	\ldots	&	\ldots	&	\ldots	&	1.4708	&	\ldots	&	\ldots	&	\ldots	&	n	&	Strong EB in K2 data, no $P$ at 3 d.	\\
248023915	&	 2MASS J04380083+2558572 	&	\ldots	&	\ldots	&	\ldots	&	  	\ldots	&	\ldots	&	\ldots	&	2	&	\ldots	&	0.6644	&	1.0263	&	\ldots	&	\ldots	&	n	&	No period near 2 d in K2 data.	\\
248029373	&	 DKTauAB 	&	8.4	&	\ldots	&	8.18	&	  	\ldots	&	4.094	&	\ldots	&	\ldots	&	\ldots	&	7.8414	&	\ldots	&	\ldots	&	\ldots	&	n	&	Complex K2 LC, could explain mismatch. 	\\
248029954	&	 2MASS J04394748+2601407 	&	\ldots	&	\ldots	&	\ldots	&	  	\ldots	&	\ldots	&	\ldots	&	2.9	&	\ldots	&	2.9502	&	\ldots	&	\ldots	&	\ldots	&	y	&	\ldots	\\
248040905	&	 V*IQTau 	&	6.25	&	\ldots	&	\ldots	&	 	6.25	&	6.902	&	1.18	&	\ldots	&	\ldots	&	6.6721	&	\ldots	&	\ldots	&	\ldots	&	y/n	&	Rodriguez+ is the outlier.	\\
248045033	&	 TYC1833-575-1 	&	\ldots	&	\ldots	&	\ldots	&	  	\ldots	&	\ldots	&	0.952	&	\ldots	&	\ldots	&	2.6393	&	\ldots	&	\ldots	&	\ldots	&	n	&	No period near 0.9 d in K2 data.	\\
248049475	&	 V*DOTau 	&	\ldots	&	\ldots	&	\ldots	&	  	\ldots	&	\ldots	&	0.961	&	\ldots	&	\ldots	&	\ldots	&	\ldots	&	\ldots	&	\ldots	&	n	&	Rigon etal\ (2017) reports  5-6d, 24d, 6-7d, 17d, 35d. No period in K2, but LC has a lot of structure, suggesting that periodic behavior at another time may be plausible.	\\
248055184	&	 ITTauA 	&	\ldots	&	\ldots	&	\ldots	&	  	\ldots	&	\ldots	&	2.751	&	\ldots	&	\ldots	&	2.7434	&	\ldots	&	\ldots	&	\ldots	&	y	&	\ldots	\\
248060724	&	 [BLH2002]KPNO-Tau14 	&	\ldots	&	\ldots	&	\ldots	&	  	\ldots	&	\ldots	&	\ldots	&	1.86	&	\ldots	&	1.8637	&	\ldots	&	\ldots	&	\ldots	&	y	&	\ldots	\\
248145565	&	 HD283798 	&	\ldots	&	0.6	&	\ldots	&	  	\ldots	&	\ldots	&	0.965	&	\ldots	&	0.987	&	0.9831	&	0.9658	&	\ldots	&	\ldots	&	y/n	&	Grankin+ is the outlier.	\\
248175684	&	 V*V1328Tau 	&	\ldots	&	\ldots	&	\ldots	&	  	\ldots	&	\ldots	&	0.601	&	\ldots	&	\ldots	&	1.5287	&	3.8492	&	\ldots	&	\ldots	&	n	&	No period near 0.6 d in K2 data.	\\
248180268	&	 HD283782 	&	\ldots	&	\ldots	&	\ldots	&	  	\ldots	&	\ldots	&	0.321	&	\ldots	&	0.87	&	0.465	&	0.3214	&	2.0182	&	\ldots	&	n	&	No period near 0.8 d in K2 data; Rodriguez+ matches our $P_2$.	\\
248245414	&	\ldots	&	\ldots	&	0.921	&	\ldots	&	  	\ldots	&	\ldots	&	\ldots	&	\ldots	&	\ldots	&	\ldots	&	\ldots	&	\ldots	&	\ldots	&	n	&	No period near 0.9 d in K2 data.	\\
248248731	&	\ldots	&	\ldots	&	2.43	&	\ldots	&	  	\ldots	&	\ldots	&	\ldots	&	\ldots	&	\ldots	&	\ldots	&	\ldots	&	\ldots	&	\ldots	&	n	&	No period near 2 d in K2 data.	\\
\enddata
\end{deluxetable}
\end{longrotatetable}

\clearpage

\end{document}